\documentclass[article]{rmaa}

\newcommand{\elecd}{$n_{\rm e}$}
\newcommand{\te}{$T_{\rm e}$}
\newcommand{\hb}{H$\beta$}
\newcommand{\ha}{H$\alpha$}

\newcommand{\foiii}{[\ion{O}{3}]}

\newcommand{\foii}{[\ion{O}{2}]}
\newcommand{\fsii}{[\ion{S}{2}]}
\newcommand{\fsiii}{[\ion{S}{3}]}

\newcommand{\fnii}{[\ion{N}{2}]}

\newcommand{\ffeii}{[\ion{Fe}{2}]}
\newcommand{\ffeiii}{[\ion{Fe}{3}]}

\newcommand{\nitroi}{\ion{N}{1}}
\newcommand{\nii}{\ion{N}{2}}

\newcommand{\oi}{\ion{O}{1}}
\newcommand{\oii}{\ion{O}{2}}

\newcommand{\cii}{\ion{C}{2}}

\newcommand{\sii}{\ion{S}{2}}
\newcommand{\siii}{\ion{S}{3}}

\newcommand{\cliii}{\ion{Cl}{3}}
\newcommand{\fariii}{[\ion{Ar}{3}]}

\newcommand{\feiii}{\ion{Fe}{3}}

\newcommand{\hi}{\ion{H}{1}}
\newcommand{\hii}{\ion{H}{2}}
\newcommand{\di}{\ion{D}{1}}
\newcommand{\hei}{\ion{He}{1}}

\newcommand{\ts}{\emph{$t^2$}}

\SetYear{2006}

\title{The chemical composition of the Galactic \ion{H}{2} regions M8 and M17. A revision based on
deep VLT echelle spectrophotometry\footnotemark{}}

\author{J. Garc\'{\i}a-Rojas,\altaffilmark{1}
	C. Esteban,\altaffilmark{1}
	A. Peimbert,\altaffilmark{2}
	M. Rodr\'{\i}guez,\altaffilmark{3}
	M. Peimbert,\altaffilmark{2}
	and M.~T. Ruiz\altaffilmark{4}} 
	
\altaffiltext{1}{Instituto de Astrof{\'\i}sica de Canarias, Spain}

\altaffiltext{2}{Instituto de Astronom\'\i a, UNAM, M\'exico}

\altaffiltext{3}{Instituto Nacional de Astrof\'{\i}sica, \'Optica y Electr\'onica, Mexico}

\altaffiltext{4}{Departamento de Astronom\'{\i}a, Universidad de Chile, Chile}

\fulladdresses{
\item Jorge Garc\'{\i}a-Rojas and C\'esar Esteban: Instituto de Astrof\'{\i}sica de Canarias, 
V\'{\i}a L\'actea s/n, E-38200, La Laguna, Tenerife, Spain (jogarcia@iac.es, cel@iac.es).
\item Antonio Peimbert and Manuel Peimbert: Instituto de Astronom\'\i a, UNAM, 
Apdo. Postal 70-264, 04510 M\'exico, D.F., Mexico (antonio@astroscu.unam.mx, peimbert@astroscu.unam.mx).
\item M\'onica Rodr\'{\i}guez: Instituto Nacional de Astrof\'{\i}sica, \'Optica y Electr\'onica, Apdo. Postal 51 y 
216,  
72000 Puebla, Mexico (mrodri@inaoep.mx).
\item Maria Teresa Ruiz: Departamento de Astronom\'{\i}a, Universidad de Chile,
Casilla Postal 36D, Santiago de Chile, Chile (mtruiz@das.uchile.cl).}

\shortauthor{Garc\'{\i}a-Rojas, Esteban, Peimbert et al.}
\shorttitle{ The \ion{H}{2} regions M8 and M17}

\abstract{We present new echelle spectrophotometry of the Galactic \ion{H}{2} regions M8 and M17. The data have 
been taken 
with the VLT UVES echelle spectrograph in the 3100 to 10400 {\rm \AA}\ range.
We have measured the intensities of 375 and 260 emission lines in M8 and M17 respectively, increasing 
significatively the number of emission lines measured in previous spectrophotometric studies of these nebulae. 
Most of the detected lines are permitted lines. Electron 
temperatures and densities have been determined using different diagnostics.
We have derived He$^{+}$, C$^{++}$, O$^{+}$ and O$^{++}$ ionic abundances from pure 
recombination lines. We have also derived 
abundances from collisionally excited lines for 
a large number of ions of different elements. Highly consistent 
estimations of {\ts} have been obtained by using different independent indicators, the values are 
moderate and very similar to those obtained in other 
Galactic {\hii} regions.
We report the detection of deuterium Balmer 
emission lines, up to D$\epsilon$, in M8 and show that their intensities are consistent with 
continuum fluorescence as their main excitation mechanism. 
}

\resumen{Presentamos nuevos datos espectrofotom\'etricos de las regiones {\hii} Gal\'acticas M8 y M17. Los 
datos han sido obtenidos usando el espectr\'ografo echelle UVES del VLT en el rango entre los 3100 y los 10400 {\rm 
\AA}. 
Hemos medido las intensidades de 375 y 260 l\'{\i}neas de emisi\'on en M8 y M17, respectivamente, incrementando de 
forma 
significativa el n\'umero de l\'{\i}neas identificadas en estas 
nebulosas. La mayor\'{\i}a de las l\'{\i}neas detectadas son permitidas. Se han calculado las temperaturas 
y densidades electr\'onicas usando diferentes diagn\'osticos. Hemos determinado las abundancias i\'onicas de 
He$^{+}$, C$^{++}$, O$^{+}$ and O$^{++}$ a partir de l\'{\i}neas debidas \'unicamente a recombinaci\'on. Tambi\'en 
hemos determinado 
las abundancias de un gran n\'umero de iones de diferentes elementos usando l\'{\i}neas de excitaci\'on colisional. 
Se han obtenido estimaciones consistentes de {\ts} usando diferentes indicadores independientes. 
Detectamos l\'{\i}neas de emisi\'on de la serie de Balmer de deuterio en M8, hasta D$\epsilon$; 
tambi\'en mostramos que sus intensidades son consistentes con el hecho de que la fluorescencia del cont\'{\i}nuo 
sea el 
principal mecanismo de excitaci\'on de estas l\'{\i}neas. 
}

\addkeyword{line: identification}
\addkeyword{ISM: abundances---H II regions}
\addkeyword{individual: M8, M17}


\begin{document}

\maketitle

\section{Introduction}
\label{intro}

\footnotetext{Based on observations collected at the European Southern
Observatory, Chile, proposal number ESO 68.C-0149(A).}

This is the last of a series of 5 papers devoted to present results of long-exposure high-spectral-resolution  
spectral 
data taken with the VLT UVES echelle spectrograph with the aim of obtaining accurate  
measurements of very 
faint permitted lines of heavy element ions in Galactic {\hii} regions. Our sample consists of   
eight of the brightest
Galactic {\hii} regions which cover a range of Galactocentric distances from 6.3 to 10.4 kpc  
(assuming the Sun 
to be at 8 kpc from the Galactic center). The objects whose data have already been published are: NGC~3576  
\citep{garciarojasetal04}, 
the Orion Nebula \citep{estebanetal04}, S~311 \citep{garciarojasetal05}, M16, M20, and NGC~3603 
\citep{garciarojasetal06}. 

Along this project we have detected and measured an unprecedented large number of emission  
lines in all the {\hii} regions analyzed, which could improve the knowledge of the nebular gas conditions 
and abundances. 
We have derived chemical abundances of C$^{++}$ and O$^{++}$ from several recombination 
lines of {\cii} and {\oii}, avoiding the problem of line blending in all the {\hii} regions of our sample. 
The high signal-to-noise ratio of the VLT spectra of M8 and M17 has allowed us to detect and measure more 
C$^{++}$ and O$^{++}$ RLs than in previous works (i.e., Esteban et al. 1999b, hereinafter EPTGR, in M8 and Esteban 
et al. 1999a, 
hereinafter EPTG, in M17); 
also, the reliability of these lines 
has increased significantly with respect to the previous detections. 
{From} the observations of all the objects of our project, \citet{estebanetal05} obtained --for the first time-- 
the radial 
gas-phase C and O gradients of the Galactic disk making use of RLs, which are, in principle, better for abundance 
determinations because the ratio $X^{+p}$/H$^+$ from RLs is almost-independent of the temperature structure of the 
nebula. 
A reliable determination of these gradients is of paramount importance for chemical evolution models of our Galaxy 
\citep[see][]{carigietal05}.

The fact that ionic abundances determined from the intensity of collisionally excited lines (CELs) are 
systematically  
lower (with factors ranging from 1.3 to 2.8) than those 
determined by recombination lines (RLs) is far from being completely understood, and has led to the so-called 
``abundance discrepancy'' problem. This problem is clearly present in Galactic {\hii} regions 
\citep[see][and all papers related to this project]{peimbertetal93b, tsamisetal03}.
In the case of extragalactic studies, only a few works have been developed with the aim of  
detecting the 
faint recombination lines: \citet{estebanetal02} for M33 and M101, \citet{apeimbert03} and  
\citet{tsamisetal03} 
for the Magellanic Clouds, \citet{apeimbertetal05} for NGC~6822. Moreover, \citet{lopezsanchezetal06}
have, for NGC~5253, estimated abundance discrepancies rather similar to those of the Galactic objects.

One of the probable causes of the abundance discrepancy is the presence of spatial variations in 
the temperature structure of the nebulae \citep{torrespeimbertetal80}. Temperature fluctuations  
may produce the discrepancy due to the different functional dependence of the line emissivities of CELs  
and RLs 
on the electron temperature, which is stronger --exponential-- in the case of CELs. 
Temperature fluctuations have been parametrized traditionally by \ts, the mean-square 
temperature fluctuation of the gas \citep[see][for a detailed formulation]{peimbert67,  
peimbertcostero69, peimbert71}. It is a well known result that photoionization codes 
cannot reproduce the temperature fluctuations found in gaseous nebulae, but there are mainly two  
possibilities to explain them: first, there might be an additional important source of energy producing such 
fluctuations, 
which has not been taken into account by photoionization models; 
second, there could be density inhomogeneities \citep{viegasclegg94} or chemical inhomogeneities 
\citep[see][and references therein]{tsamispequignot05} that produce temperature variations. 
The physical processes that may cause such temperature fluctuations 
are still subject of controversy. Reviews of the relevant processes can be found in \citet{esteban02},
\citet{torrespeimbertpeimbert03}, and \citet{peimbertpeimbert06}. 
Additionally, there are some very recent works devoted to this topic: 
e.g., \citet{giammancobeckman05} have proposed ionization 
by cosmic rays as an additional source of energy to reproduce the temperature fluctuations 
observed in {\hii} regions; and \citet{tsamispequignot05} have developed photoionization 
models for 30 Doradus in the Large Magellanic Cloud that reproduce observed temperature fluctuations through 
chemical 
inhomogeneities (inclusions) due to the infall of material nucleosynthetically processed in supernova 
events. Further studies are needed to understand this problem.

Several spectrophotometric works devoted on the chemical composition of M8 and M17 have
been carried out previously. For M8, there are several low and intermediate spectral resolution studies
\citep{rubin69, peimbertcostero69, sanchezpeimbert91, peimbertetal93b, rodriguez99b} and one 
high spectral resolution study (EPTGR). 
The chemical abundances of M17 have been studied using low resolution spectroscopy \citep{rubin69, 
peimbertcostero69, 
peimbertetal92, rodriguez99b, tsamisetal03} and high-spectral resolution data (EPTG).

In this paper we make a reappraisal of the chemical composition of M8 and M17 in the  
same slit position observed by EPTGR in M8 and one of the positions observed by EPTG in M17 (position 14), 
by means of new echelle spectrophotometry obtained with the ESO's Very Large Telescope. Our new 
observations increase significantly the number of lines detected and the quality of the measured line intensities 
for these two nebulae.

In \S\S~\ref{obsred} and~\ref{lin} we describe the observations, the data reduction, and line intensity 
determination procedures. 
In \S~\ref{phiscond} we obtain temperatures and densities 
using several diagnostic ratios.
In \S~\ref{helioabund} we briefly analyze the recombination spectra of  
{\hei} and derive the He$^{+}$/H$^{+}$ ratio. In \S~\ref{cels} we give the ionic abundances 
determined from CELs. In \S~\ref{recom} we use RLs to derive O and C ionic abundances. 
In \S~\ref{abuntot} we present the total abundances. We report the detection of deuterium Balmer lines  
in \S~\ref{deuterium}. In \S\S~\ref{comp} 
and ~\ref{conclu} we present the comparison with previous results and the conclusions, respectively.

\section{Observations and Data Reduction}
\label{obsred}

The observations were made on 2002 March 11 with the Ultraviolet Visual Echelle Spectrograph, UVES 
\citep{dodoricoetal00}, 
at the VLT Kueyen Telescope in Cerro Paranal Observatory (Chile). We used the standard settings in 
both the red and 
blue arms of the spectrograph, covering the region from 3100 to 10400 {\rm \AA}\ . The log of the 
observations is presented in
Table~\ref{tobs}.

\setcounter{table}{0}
\begin{table}[htbp]
\centering 
\setlength{\tabnotewidth}{\columnwidth}
\tablecols{3}
\setlength{\tabcolsep}{2.8\tabcolsep}
\scriptsize
\caption{Journal of observations.}
\label{tobs}
\begin{tabular}{l@{\hspace{2.8mm}}c@{\hspace{2.8mm}}c@{\hspace{2.8mm}}}
\toprule
& \multicolumn{2}{c}{Exp. time (s)}\\
\cmidrule{2-3}
$\Delta\lambda$~(\AA) & M8 & M17 \\
\midrule
3000--3900 & 30, 3 $\times$ 300 & 30, 3 $\times$ 300 \\
3800--5000 & 30, 3 $\times$ 800 & 60, 3 $\times$ 800 \\
4700--6400 & 30, 3 $\times$ 300 & 30, 3 $\times$ 300  \\
6300--10400& 30, 3 $\times$ 800 & 60, 3 $\times$ 800  \\
\midrule
\end{tabular}
\end{table}

The wavelength regions 5783--5830 {\rm \AA}\ and 8540--8650
{\rm \AA}\ were not observed due to a gap between the two CCDs used in
the red arm. There are also five small gaps that were not observed, 9608--9612 {\rm \AA}, 
9761--9767 {\rm \AA}, 9918--9927 {\rm \AA}, 
10080--10093 {\rm \AA}\ and 10249--10264 {\rm \AA}, because  
the five redmost orders did not fit completely within the CCD.  We took long and short exposure 
spectra to check for possible saturation effects.

The slit was oriented east-west and the
atmospheric dispersion corrector (ADC) was used to keep the same observed
region within the slit regardless of the air mass value (the averaged $\sec$ z are $\sim$ 1.4 for M17 and $\sim$ 
1.85 for M8).  
The slit width was
set to 3.0$\arcsec$ and the slit length was set to 10$\arcsec$ in the blue arm and to 12$\arcsec$ in
the red arm; the slit width was chosen to maximize the S/N ratio of the
emission lines and to maintain the required resolution to separate most of the
weak lines needed for this project. The effective resolution for the lines
at a given wavelength is approximately $\Delta \lambda \sim \lambda / 8800$. 
The center of the slit was located in the same position than in EPTGR for M8 (labeled as HGS) 
and is coincident with position 14 of EPTG for M17. The final 1D spectra were extracted from an area  
of 3$\arcsec$$\times$8.3$\arcsec$. 
 
The spectra were reduced using the {\sc IRAF}\footnotemark{} echelle reduction
package, following the standard procedure of bias subtraction, 
flatfielding, aperture extraction, wavelength calibration and flux calibration. 
The standard star EG~247 was observed for flux calibration. We have not attempted sky subtraction 
from the spectra due to the slit length is much smaller than the objects; also, the spectral resolution of our data 
permit us to clearly distinguish among the telluric lines and the nebular ones.

\footnotetext{{\sc IRAF} is distributed by NOAO, which is operated by AURA,
under cooperative agreement with NSF.}

\section{Line Intensities and Reddening Correction}
\label{lin}

Line intensities were measured integrating all the flux in the line between two 
given limits and over a local continuum estimated by eye. In the cases of line blending, 
a multiple Gaussian profile fit procedure was applied to obtain the line flux of each 
individual line. Most of these measurements were made with the SPLOT routine of the {\sc IRAF} 
package. In some cases of very tight blends or blends with very bright telluric lines the 
analysis was performed via Gaussian fitting making use of the Starlink DIPSO software 
\citep{howardmurray90}.

Table~\ref{lineidm8m17} presents the emission line intensities of M8 and M17, 
respectively. The first and fourth columns include the adopted laboratory wavelength, $\lambda_0$,  
and 
the observed wavelength in the heliocentric framework, $\lambda$. 
The second and third columns show the ion and the multiplet number, or 
series for each line.  The fifth and sixth columns list the observed
flux relative to H$\beta$, $F(\lambda$), and the flux corrected for reddening
relative to H$\beta$, $I(\lambda$). The seventh column includes the
fractional error (1$\sigma$) in the line intensities relative to H$\beta$, $I(\lambda$). 
Errors were derived following 
\citet{garciarojasetal04}, adding quadratically the error due to flux calibration, 
that has been assumed as 3\%, as estimated in \citet{garciarojasetal06}, for 
similar data taken with the same instrumentation, and for which there were 
additional standard stars. Fractional error in the line fluxes relative to H$\beta$, $F(\lambda$), can be estimated 
taking into account that fractional errors in column seven were computed propagating the uncertainty in the 
extinction 
correction.

A total of 375 and 260 emission lines were measured in M8 and M17, respectively. Most of the lines are  
permitted. We 
have measured 97 forbidden lines in M8, and 52 in M17. We have detected also 5 semiforbidden lines  
in M8.
Several lines were strongly affected by atmospheric absorption features or by 
charge transfer in the CCD, rendering their intensities unreliable. Also, some  
lines 
are dubious identifications and 3 emission lines in M8 could not be identified in any of the  
available
references. All those lines are indicated in Table~\ref{lineidm8m17}.

The identification and adopted laboratory wavelengths of the lines were obtained following 
previous identifications in the literature \citep[see][and references  
therein]{estebanetal04,garciarojasetal04}. Several previously unidentified lines in M8 (EPTGR)
have been identified (see Table~\ref{lineidm8m17}).
Lines unidentified by EPTGR in M8 which are not in our line list are probably  
telluric emission lines or nebular lines which were severely blended with telluric lines. In particular, 
the features at 5865.15 {\rm \AA}, 6863.45 {\rm \AA}, and 8833.17 {\rm \AA} were identified by 
\citet{osterbrocketal96} as OH night-sky lines. 
We have identified $\lambda$10021.05 {\rm \AA} line as a telluric line. Also,  
the two lines not identified by EPTG in position 14 of M17 have been identified here as {\hei} 
$\lambda\lambda$7160.58 (1/10), 
8486 (6/16). 

It is known that the main ionization souce for the hourglass nebula in M8 is the star H36, and that 
it shows a considerably higher extinction that other zones of M8. For H36, the $A_v/E(B-V)$ ratio, $R$, has been 
determined as 4.6 by 
\citet{hechtetal82} and as 5.3 by \citet{cardellietal89}. 
Following \citet{sanchezpeimbert91} and \citet{peimbertetal93b} we have adopted for this zone of M8 a reddening  
function with R$_v$ = 5.0 parametrized by \citet{cardellietal89} for $\lambda$ $\geq$ 4100  {\rm \AA}.
A reddening coefficient of c({\hb}) = 0.94  
$\pm$ 0.03 was derived. This value is intermediate between c({\hb}) = 0.85 $\pm$ 0.05 obtained by EPTGR and  
c({\hb}) = 1.00 $\pm$ 0.10 derived by \citet{sanchezpeimbert91} and \citet{peimbertetal93b} for the same slit  
position. For the reddening function assumed for $\lambda$ $<$ 4100  {\rm \AA} see \S~\ref{extalt}.

For M17, we have adopted the standard extinction for the Milky Way parametrized by  
\citet{seaton79}. 
We have obtained a reddening coefficient of c({\hb}) = 1.17 $\pm$ 0.05, which is also intermediate  
between 
the values obtained by EPTG (1.05 $\pm$ 0.05) and \citet{peimbertetal92} (1.20) for the same slit  
position.

\subsection{Extinction correction in M8 for $\lambda$ $<$ 4100  {\rm \AA}.}
\label{extalt}

\begin{figure}[htbp]
\includegraphics[width=\columnwidth]{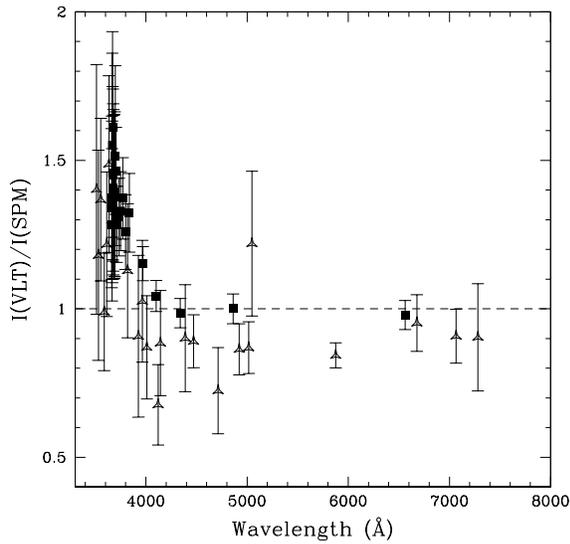}
\caption{Line flux ratio of {\hi} Balmer lines (squares) and {\hei} lines (triangles) measured in this work with
respect to those measured by EPTGR for M8 (see text).}
\label{vltspm}
\end{figure}

In Figure~\ref{vltspm} we show the ratio of the observed fluxes of {\hi} Balmer lines and {\hei} lines 
measured by us and by EPTGR. It can be seen that for wavelengths shorter than 
4100 {\rm \AA} our line fluxes are higher than those measured by EPTGR. So if we assume the  
extinction correction adopted above, the intensity of these lines would be overestimated. 
The effect seems to be an observational bias instead of a physical effect; actually M8 is the object 
that was observed at the highest air mass --$\sec$ z $\sim $ 2--, so it is possible that 
an unsuitable operation of the Atmospheric Dispersion Corrector (ADC) at high airmasses used on our 
observations caused this effect. The gradient in the reddening and in the surface brightness of the 
Hourglass region is very strong and atmosphere refraction effects could include regions of higher 
emissivity in the blue part of the spectrum that are not included at $\lambda$ $>$ 4100 \AA.

\begin{figure}[htbp]
\includegraphics[width=\columnwidth]{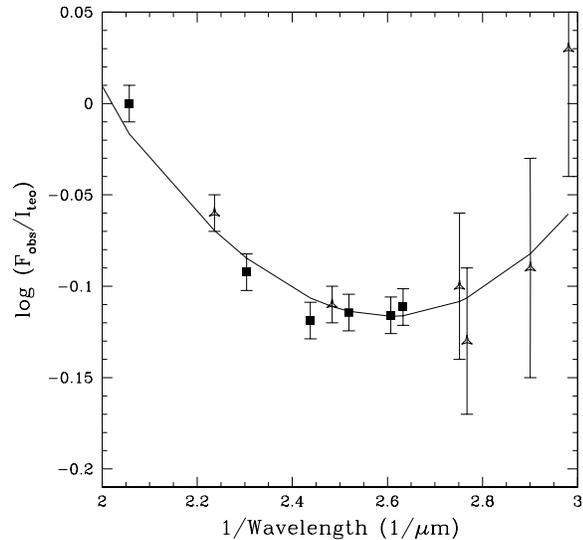}
\caption{Polynomial fit to the ratio of observed over theoretical fluxes of some {\hi} Balmer lines 
(from H10 to H$\beta$, squares) and some {\hei} lines (triangles). Note that for $\lambda$ $<$ 
4100 {\rm \AA} (1/$\lambda$ $>$ 2.44) the behavior of the lines is anomalous (see text).}
\label{extm8}
\end{figure}

To correct for this effect, we have done a polynomial fit to the observed over theoretical flux ratios of {\hi} 
Balmer lines and {\hei} 
lines which are in case B and are not affected by self-absorption effects: {\hei} $\lambda\lambda$ 3354.55, 
3447.59, 3613.64, 
3634.25, 4026.08, and 4471.48, 
and interpolated to all wavelengths shortwards of 4100 {\rm \AA} (see Figure~\ref{extm8}). 
We have not included {\hi} Balmer lines with quantum number higher than 10 in this fit due to the higher 
dependence of these line ratios with density \citep[see e.g.,][]{zhangetal04}. 
This fit is used to interpolate for all the wavelengths shortwards of 4100 {\rm \AA}. The correction has not 
affected 
significantly the physical conditions and the chemical abundances derived in this work --less than 0.05 and 0.1 dex 
in the total 
abundances of O and Ne, respectively, which are the most affected species by this effect--. 

\section{Physical Conditions}
\label{phiscond}

\subsection{Temperatures and Densities}
\label{temden}

We have derived  physical conditions  of the two nebulae using several emission line ratios. The 
temperatures and 
densities are presented in Table~\ref{plasma}. The values of {\te} and {\elecd}  
were derived using the {\sc IRAF} task \emph{temden} of the package \emph{nebular} \citep{shawdufour95} with 
updated  
atomic data 
\citep[see][]{garciarojasetal05}, except in the case of the {\elecd} 
derived from {\ffeiii} lines. We have derived the {\ffeiii} density from the intensity of the  
brightest lines --those with 
errors smaller than 30 \% and that do not seem to be affected by line blending, which are 14 in the case of 
M8 and 5 in the case of M17-- together with the computations of \citet{rodriguez02}, following the 
procedure described by \citet{garciarojasetal06}.

We have derived a weighted mean of {\elecd}({\oii}), {\elecd}({\sii}), {\elecd}({\cliii}), and  
{\elecd}({\feiii}) 
assuming an initial temperature of {\te} = 10000 K, then we have used this density to compute the  
temperatures, and 
iterated until convergence. The adopted {\elecd} values are shown in  
Table~\ref{plasma}. We have excluded 
{\elecd}({\nitroi}) from the average because this ion is representative of the very outer part of  
the nebula, and it does not coexist with the other ions.

Electron temperatures from forbidden lines have been derived from {\foii}, {\foiii}, {\fnii},  
{\fsii}, {\fsiii}, and {\fariii} line ratios. 

\begin{table}[htbp]\centering
\setlength{\tabnotewidth}{\columnwidth}
\tablecols{4}
\scriptsize
\caption{Plasma Diagnostic.}
\label{plasma}
\begin{tabular}{llll}
\toprule
 &  & \multicolumn{2}{c}{Value} \\
\cmidrule{3-4}
Parameter & Line & \multicolumn{1}{c}{M8}& \multicolumn{1}{c}{M17} \\
\midrule
N$_{\rm e}$ (cm$^{-3}$)& [N\thinspace I] & 1600$^{+750}_{-470}$ &  
1200$^{+1250}_{-500}$  \\
& [O\thinspace II] & 1800 $\pm$ 800& 480 $\pm$ 150  \\
& [S\thinspace II] & 1600$\pm$450& 500$\pm$220 \\ 
& [Fe III] &  2600$\pm$1450 &  430$^{+>1000}_{-400}$ \\
& [Cl\thinspace III] & 2100$\pm$700 & 270$^{+630}_{-270}$ \\ 
& [Ar\thinspace IV] & 2450: & $>$800 \\ 
& N$_{\rm e}$ (adopted) & 1800$\pm$350 & 470$\pm$120  \\
& & & \\
T$_{\rm e}$ (K)& [N\thinspace II] & 8470$\pm$ 180\tabnotemark{a} & 8950$\pm$ 380\tabnotemark{a} \\
& [S\thinspace II] &  7220$\pm$300 & 7100$\pm$750  \\
& [O\thinspace II] & 8700$\pm$350\tabnotemark{a} & 8750$\pm$550\tabnotemark{a} \\
& T$_{\rm e}$ (low) & 8500 $\pm$ 150 & 8870$\pm$300 \\
& [O\thinspace III]  & 8090$\pm$ 140& 8020$\pm$ 170  \\
& [Ar\thinspace III] & 7550$\pm$420 & 8380$\pm$570  \\
& [S\thinspace III]  & 8600$\pm$300\tabnotemark{b} & 8110$\pm$400  \\
& T$_{\rm e}$ (high) & 8150$\pm$120 & 8050$\pm$150 \\
& He\thinspace I &  7650$\pm$200 & 7450$\pm$200 \\
& Balmer line/cont. & 7100$^{+1250}_{-1000}$ & \nodata \\ 
& Paschen line/cont. & 7750$\pm$900 & 6500$\pm$1000 \\ 
\bottomrule
\tabnotetext{a}{Recombination contribution on the auroral lines has been considered (see text)}
\tabnotetext{b}{{\fsiii} $\lambda$9530 affected by atmospheric absorption bands.}
\end{tabular}
\end{table}

We have corrected {\te}({\oii}) from the contribution to $\lambda\lambda$7320+7330 due to 
recombination following the formula derived by \citet{liuetal00}:
\begin{equation}
\frac{I_R(7320+7330)}{I({\rm H\beta})}
= 9.36\times(T_4)^{0.44} \times \frac{{\rm{O}}^{++}}{{\rm{H}}^+},
\end{equation}
where $T_4$=$T$/10$^4$. Using the O$^{++}$/H$^+$ ratio derived 
by EPTGR in M8 and EPTG in M17 from RLs we have estimated contributions of about 2\% and 20\% for M8  
and M17, respectively. 
The large contribution of recombination to the intensity of the $\lambda\lambda$7320+7330 lines in  
M17 is reflected in a drop of more than 1000 K in {\te}({\oii}), which reconciles the value of {\te}({\oii}) 
with that of {\te}({\nii}).

\citet{liuetal00} also give a formula for the contribution by recombination to the intensity of the {\fnii} 
$\lambda$ 5755 
line:
\begin{equation}
\frac{I_R(5755)}{I({\rm H\beta})}
= 3.19\times(T_4)^{0.30} \times \frac{{\rm N}^{++}}{{\rm H}^+}.
\end{equation}
To derive the N$^{++}$/H$^+$ ratio, needed to 
compute this quantity, we have assumed that N$^{++}$/H$^+$ is well represented by the subtraction of  
N$^+$/H$^+$ to the total N/H ratio, 
assuming that the temperature fluctuations paradigm and a ionization correction factor (hereinafter ICF)
leads to the correct abundances 
(see \S~\ref{tempvar}).\footnote{Another way to derive the  
N$^{++}$/H$^+$ ratio is 
assuming as valid the abundance obtained from {\nii} lines of multiplet 3, which seems to be the  
least affected by fluorescence effects. Nonetheless, for regions with high degree of ionization, it may be 
incorrect 
to apply permitted line abundances because as pointed out by \citet{grandi76}, {\nii} permitted lines are excited 
mainly by resonance fluorescence, and corrections might be high. In fact, if we assume the N$^{++}$ abundance 
derived 
from multiplet 3, the correction would be of more than 20\%, implying a {\te}({\fnii}) 500 K lower than that has 
been assumed (see \S~\ref{recom} for additional discussion on the {\nii} permitted lines).}
{From} the results of EPTGR for M8 and EPTG for M17, 
the contribution of recombination to the intensity of the {\fnii} $\lambda$5755 line has been  
estimated as 1 \% and 6 \% for M8 and M17, respectively. These contributions are small and 
affect in less than 200 K the derived temperature.

\begin{figure}[htbp]
\includegraphics[width=0.85\columnwidth]{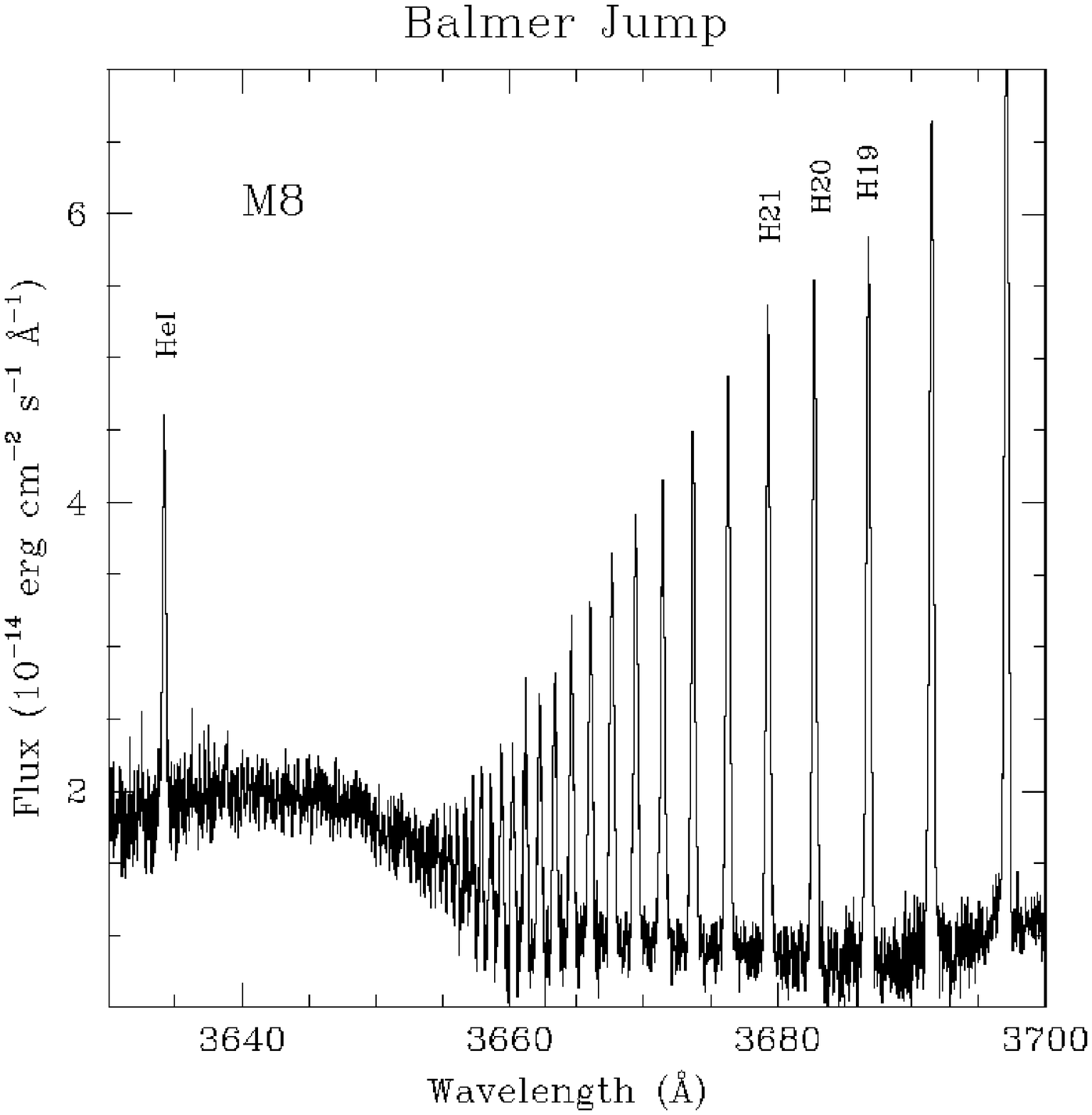}
\includegraphics[width=0.85\columnwidth]{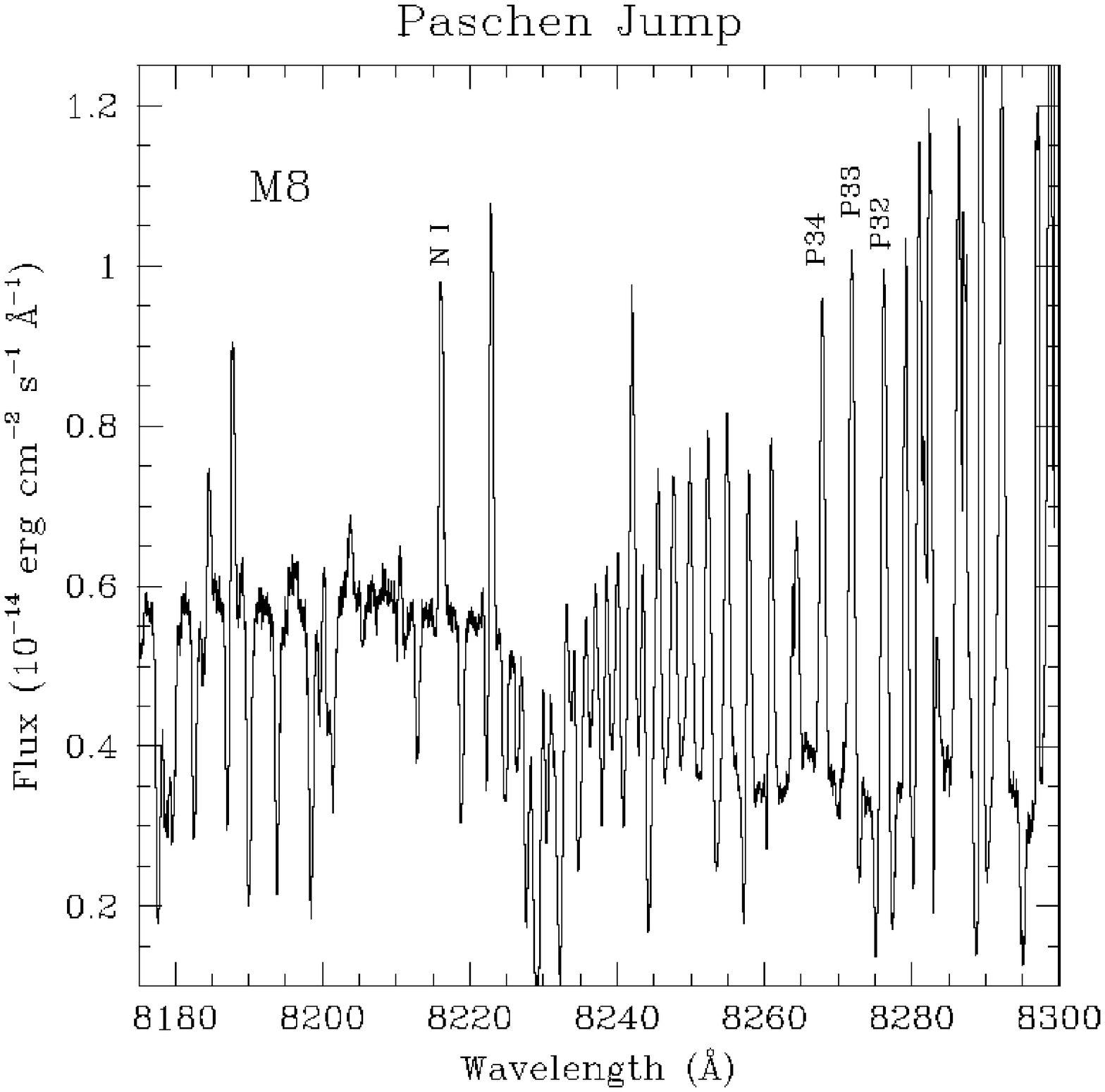}
\includegraphics[width=0.85\columnwidth]{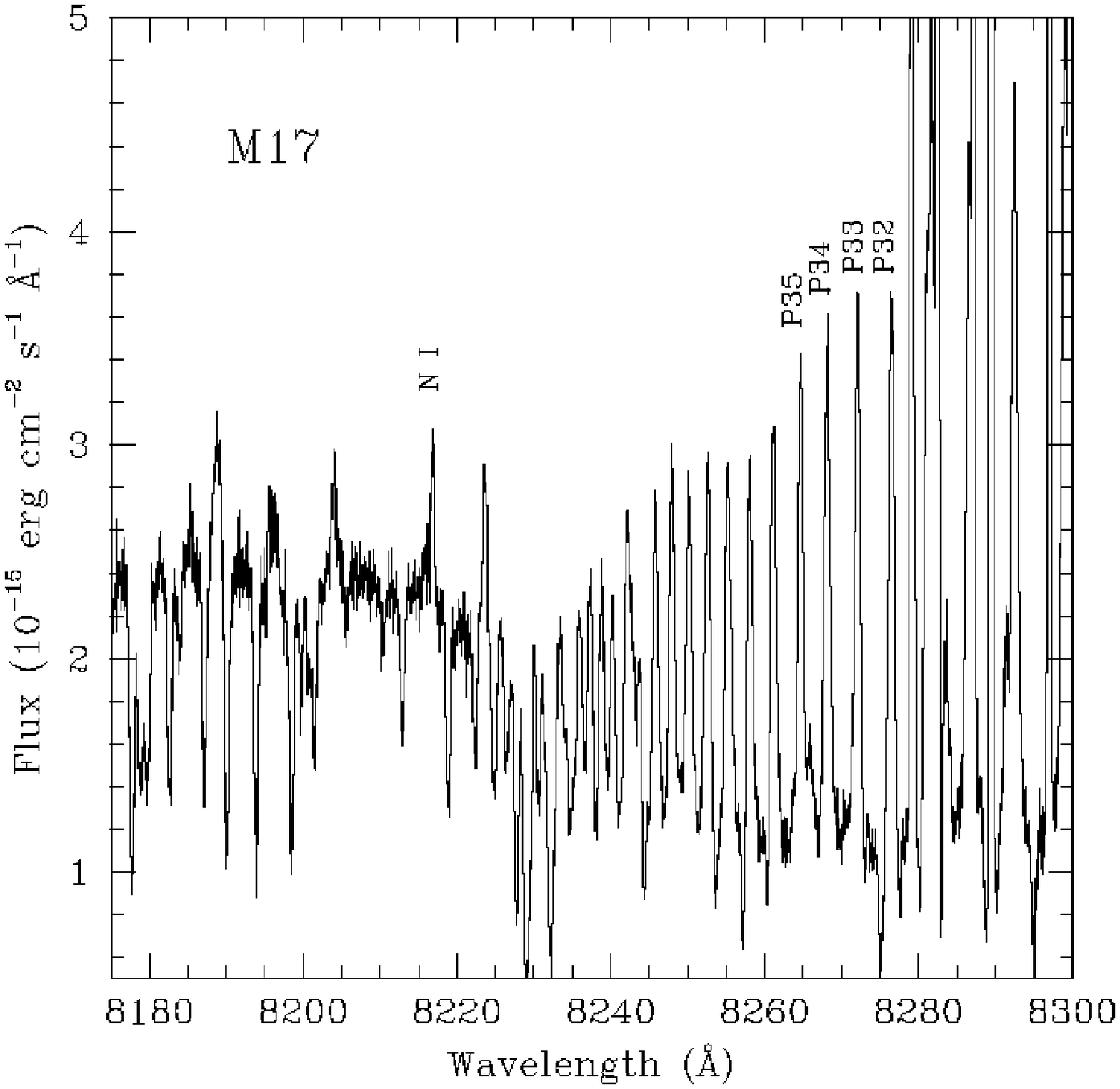}
\caption{Section of the echelle spectra of M8 (upper and middle panels) and M17 (lower panel) including the Balmer 
(upper panel) 
and the Paschen (middle and lower panels) limits (observed fluxes).}
\label{saltos}
\end{figure}

We have also been able to derive the electron temperatures from the Balmer and Paschen discontinuities.  
Figure~\ref{saltos} 
shows the spectral regions near the Balmer and the Paschen limits. The discontinuities can be  
clearly appreciated, except 
in the case of the Balmer limit of M17, for which the low signal-to-noise of the continuum makes it unreliable.  
We have followed the same 
procedure than in previous papers \citep[e.g.,][]{garciarojasetal06} to derive the temperatures. 
The values adopted for {\te}({\hi}) are shown in Table~\ref{plasma}. 
To the best of our knowledge, no previous determinations of {\te}({\hi}) (Balmer and Paschen) have been derived  
for M17; for M8 there was a previous {\te}({\hi}) determination from the Balmer discontinuity in the hourglass 
by \citet{sanchezpeimbert91} which amounts to {\te}({\hi}) = 6600 K, that is somewhat smaller than what has been 
derived here ({\te}({\hi}) = 7100$^{+1250}_{-1000}$ K), but consistent within the errors. 

Our derived {\te}({\hi}) values are in good agreement with 
the values obtained by \citet{reifensteinetal70} from the H109$\alpha$ radio recombination line, 
{\te}({\hi}) = 7300 $\pm$ 1000 K for M8 and 6400 $\pm$ 750 K for M17. On the other hand, {\te}({\hi}) derived  by 
\citet{shavergoss70} 
from radio 408 MHz continuum measurements do not agree; these authors computed {\te} = 6100 K  
for M8 and 7850 K for M17, which are far from the temperatures derived here.

We have derived {\te}({\hei}) assuming a 2-zone ionization scheme, characterized by {\te}{\sc ii+iii} 
\citep[see][]{apeimbertetal02}. 
We have derived {\te}({\hei})=7650 $\pm$ 200 K for M8, which is highly 
consistent with the {\te}({\hi}) derived above, and {\te}({\hei})=7450 $\pm$ 200 K for M17, which is  
higher than {\te}({\hi}).

We have assumed a two-zone ionization scheme for all our calculations. We have adopted the 
average of {\te} obtained from {\fnii} and {\foii} lines as representative for the low ionization  
zone. We have not included {\te}({\sii}) in the average because its value is much lower than those 
obtained from {\fnii} and {\foii} lines. This effect has been reported previously in several objects 
\citep[e.g.,][]{garciarojasetal05,garciarojasetal06}, and might be produced by the presence of a temperature 
stratification in the outer zones of the nebulae or, conversely, by errors in the atomic parameters of the ion. 
For the high ionization zone we have adopted the average of the values of {\te} obtained from {\foiii}, {\fsiii} 
and {\fariii}. In M8 the {\fsiii} $\lambda$9532 line is affected by atmospheric absorption bands, 
so we have adopted the intensity of {\fsiii} $\lambda$9069 and the {\fsiii} $\lambda$9532/$\lambda$9069 
theoretical ratio to derive {\te}({\siii}).

\subsection{Temperature Variations}
\label{tempvar}

Since \citet{torrespeimbertetal80} proposed the presence of spatial temperature fluctuations 
(parametrized by {\ts}) as a possible cause of the abundance discrepancy, many efforts have been done 
to find the physical processes responsible for such temperature fluctuations in {\hii} regions 
\citep[e.g.,][]{esteban02, tsamispequignot05} and in planetary nebulae 
\citep[e.g.,][]{liu06, peimbertpeimbert06}, but the source of temperature fluctuations and its 
impact on the chemical abundance determinations remain controversial topics in the 
study of gaseous nebulae. 

\citet{peimbert71} showed that there was a substantial difference between the {\te} derived from the {\foiii} lines 
and from the one derived from hydrogen recombination continuum discontinuities, which is strongly correlated with 
the 
abundance discrepancy \citep{liuetal01,tsamisetal04}, so the comparison between electron temperatures 
derived from both methods would be an additional indicator of {\ts}.

Additionally, it is also possible to derive the {\ts} value from the analysis of the {\hei} lines, 
because of the different temperature dependence of each of them, so we can find {\hei} line ratios 
that will allow us to derive a temperature. However, in practice, each of these ratios depends simultaneously on 
$T_0$, {\ts}, {\elecd} and $\tau_{3889}$ therefrefore, any determination must be done using several line ratios. 
\citet{peimbertetal00} developed a maximum likelihood method to search for the plasma 
conditions that would give the best simultaneous fit to the measured lines. In \S~\ref{helioabund} we have 
applied that method to our {\hei} lines.

As we have assumed a two-zone ionization scheme, we have followed the formulation of \citet{peimbertetal00} and
\citet{apeimbertetal02} to derive the values of {\ts} following the three methods described above.

In Table~\ref{t2} we show the \ts values derived from each method and the final adopted values, which are 
error-weighted averages. It is highly remarkable that all the \ts values derived for each nebula are very 
consistent. 
The C$^{++}$/H$^+$ ratio obtained from CELs for M8 has been taken from 
\citet{peimbertetal93b}, who measured the UV \ion{C}{2}] $\lambda\lambda$1906+1909 emission lines from IUE data. 
Nonetheless, as well as when we are comparing with the infrared data \citep[see][]{garciarojasetal06}, we cannot 
discard aperture effects due to the different volumes covered by the slits in the optical and UV observations.

\begin{table}[htbp]\centering
\setlength{\tabnotewidth}{\columnwidth}
\tablecols{3}
\setlength{\tabcolsep}{2.7\tabcolsep}
\scriptsize
\caption{{\ts} parameter}
\label{t2}
\begin{tabular}{ccc}
\toprule
 & \multicolumn{2}{c}{\ts} \\
\cmidrule{2-3}
Method & M8& M17 \\
\midrule
O$^{\rm ++}$ (R/C)& 0.045$\pm$0.005 & 0.034$\pm$0.005 \\
O$^{\rm +}$ (R/C)& 0.031$\pm$0.017 & 0.109:  \\
C$^{\rm ++}$ (R/C)& 0.035$\pm$0.005 & \nodata \\
He II & 0.046$\pm$0.009 & 0.027$\pm$0.014 \\
Bac/Pac--FL & 0.022$\pm$0.015 & 0.035$\pm$0.021\\ 
\midrule
Adopted & 0.040$\pm$0.004 & 0.033$\pm$0.005\\ 
\bottomrule
\end{tabular}
\end{table}

The {\ts} values obtained in this paper are very similar to those obtained for all the bright 
Galactic {\hii} regions of 
our sample \citep{estebanetal04, garciarojasetal04, garciarojasetal05, garciarojasetal06} and are also similar to 
the few 
estimations of {\ts} in extragalactic {\hii} regions available in the literature for the Magellanic 
Clouds \citep{apeimbert03, tsamisetal03}, NGC~6822 \citep{apeimbertetal05}, M101, NGC~2366, and M33 
\citep{estebanetal02} and the dwarf {\hii} galaxy NGC~5253 \citep{lopezsanchezetal06}. 

\begin{figure}[htbp]
\includegraphics[width=\columnwidth]{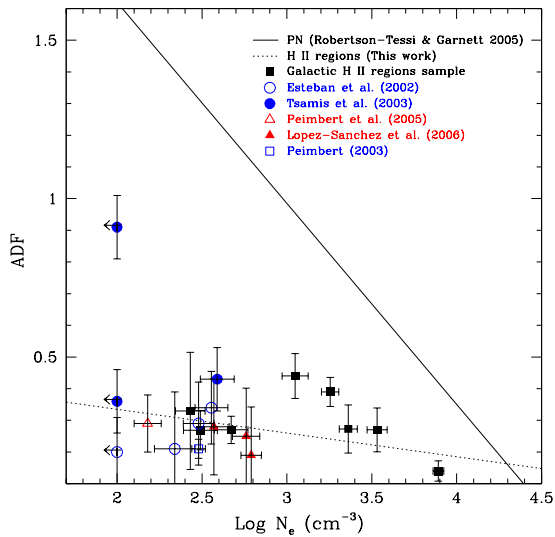}
\caption{Correlation between {\elecd} and ADF for the sample of Galactic (filled squares) and extragalactic (other 
symbols) 
{\hii} regions in which ADF has been measured. 
Solid line is the fit obtained by \citet{robertsontessigarnett05} for a sample of PN from the literature. 
Dashed line is the fit obtained for {\hii} regions. }
\label{adfne}
\end{figure}

A very different behavior has been found for planetary nebulae (PNe). Several authors have found a large scatter of 
the {\ts} values 
determined for different PNe \citep[see e.g.,][and references therein]{liu02, tsamisetal04}. Recently, 
\citet{robertsontessigarnett05}, 
have found a correlation between the {\emph Abundance Discrepancy Factor} (ADF) defined by \citet{liuetal00} as 
log(O$^{++}$/H$^+$ RL)-log(O$^{++}$/H$^+$ CEL) and {\elecd}. 
They have found that the lower {\elecd} in the PNe, the higher ADF, with a strong slope.
To illustrate the difference between both behaviors --PNe and {\hii} regions-- we have overplotted the complete 
set of ADFs measured in {\hii} regions (Galactic and extragalactic) available in the literature to the 
Robertson-Tessi's fit (r=$-$0.47) 
for PNe (see Figure~\ref{adfne}). From Figure~\ref{adfne} it is clear that {\hii} regions do not follow the 
correlation found for 
PNe. In fact, it seems that there is no correlation between the {\hii} regions data, due to the similarity between 
the ADF values found 
for {\hii} regions and to the low correlation coefficient found (r=--0.25). 
The only exception is LMC N11B, which has an ADF much larger than the other nebulae. 
For this object, \citet{tsamisetal03} corrected the intensity of the multiplet 1 {\oii} lines because of the 
presence of absorption 
features, mainly caused by the presence of B stars --which have a strong {\oii} 
absorption spectra-- on the field covered by the slit. Nevertheless, this effect,--which could be very important on 
extragalactic objects-- can only be properly corrected if the stellar absorption features are resolved, 
or if stellar population synthesis spectra are available. 
Therefore, the {\oii} absorption contribution can not be properly estimated if we have low spectral resolution data 
--like those used by Tsamis et al.-- because it is difficult 
to distinguish between emission and absorption features \citep[to illustrate this point see Figure 2 
of][]{garciarojasetal06}. It is not the 
aim of this paper to discuss more about the attenuation of the intensities of {\oii} RLs due to star absorption 
features, 
but it is important to stress that this effect should be investigated when deriving abundances from RLs in 
extragalactic {\hii} regions.

\section{He$^+$ Abundance}
\label{helioabund}

We have measured 76 and 62 {\hei} emission lines in the spectra of M8 and M17,  
respectively. 
These lines arise mainly from recombination but they can be affected by collisional excitation  
and 
self-absorption effects. We have
determined the He$^+$/H$^+$ ratio from a maximum likelihood method \citep{peimbertetal00},
using the {\elecd} of Table~\ref{plasma} and $T$(O~{\sc ii+iii})= 8350 K for M8, 
and $T$(O{\sc ii+iii})= 8200 K for M17  (see \S~\ref{tempvar}).
We have used the effective recombination coefficients of \citet{storeyhummer95} 
for {\hi} and those of \citet{smits96} and \citet{benjaminetal99} for {\hei}. 
The collisional contribution
was estimated from the calculations made by \citet{saweyberrington93} and \citet{kingdonferland95}, and the optical 
depths 
in the triplet lines were derived from the computations by \citet{benjaminetal02}. 

\begin{table}[htbp]\centering
\setlength{\tabnotewidth}{\columnwidth}
\tablecols{3}
\setlength{\tabcolsep}{2.8\tabcolsep}
\scriptsize
\caption{He$^+$ abundance.}
\label{abhe}
\begin{tabular}{lcc}
\toprule
& \multicolumn{2}{c}{He$^+$/H$^+$ \tabnotemark{a}} \\
\cmidrule{2-3}
Line & \multicolumn{1}{c}{M8}& \multicolumn{1}{c}{M17} \\
\midrule
3819.61& 673 $\pm$ 54&  950 $\pm$ 95 \\
3964.73& 733 $\pm$ 59&  897 $\pm$ 45 \\
4026.21& 699 $\pm$ 56&  955 $\pm$ 38 \\
4387.93& 679 $\pm$ 20&  918 $\pm$ 55 \\
4471.09& 662 $\pm$ 20&  904 $\pm$ 27 \\
4713.14& 617 $\pm$ 19&  952 $\pm$ 57 \\
4921.93& 670 $\pm$ 20&  896 $\pm$ 27 \\
5875.64& 639 $\pm$ 19&  880 $\pm$ 26 \\
6678.15& 666 $\pm$ 20&  924 $\pm$ 37 \\
7065.28& 673 $\pm$ 20&  905 $\pm$ 45 \\
7281.35& 666 $\pm$ 20&  884 $\pm$ 44 \\ 
\midrule
Adopted\tabnotemark{b}& 662 $\pm$ 9   & 910 $\pm$ 14  \\ 
\bottomrule
\tabnotetext{a}{In units of 10$^{-4}$, for $\tau_{3889}$ = 8.28 $\pm$ 0.60 and 7.80 $\pm$ 0.78, and $t^2$ = 0.040 
$\pm$ 0.004 and 0.033 $\pm$ 0.005, respectively. 
Uncertainties correspond to line intensity errors.}
\tabnotetext{b}{It includes all the relevant uncertainties in emission line intensities, $n_e$, $\tau_{3889}$ and 
$t^2$.}
\end{tabular}
\end{table}

In Table~\ref{abhe} we have included the He$^+$/H$^+$ ratios that we have obtained for the individual 
{\hei} lines not affected by line blending and with the highest signal-to-noise ratio, 
excluding {\hei} $\lambda$5015 because it could suffer self-absorption effects from the 2$^1$S metastable 
level, and  
$\lambda$3889 because it is severely blended with the H8 line. 
We have done a $\chi^2$ 
optimization of the values in the table, and we have obtained a $\chi^2$ parameter of 8.5 for M8 and 
3.6 for M17. The values obtained indicate that the fits are good for a system with 
eight degrees of freedom. 
EPTGR, who covered a region slightly larger than ours, but centered in the same location,
derived a He$^{+}$/H$^+$ ratio for M8--HG a factor of 1.13 (0.05 dex) higher 
than ours. On the other hand, \citet{peimbertetal93b}, who covered a similar region of the 
nebula, obtained a He$^{+}$/H$^+$ ratio 0.025 dex lower, 
confirming the strong variation with position of the values found for the He$^{+}$/H$^+$ fraction in the Hourglass 
on M8.
For M17, the He$^{+}$/H$^+$ ratio is only a bit smaller than those obtained by EPTG 
(0.0975) and \citet{peimbertetal92} (0.099); this could be due to differences on the ionization degree 
of the regions covered by the different slit sizes since the O$^{++}$ abundance from RLs shows a similar 
behavior (see \S~\ref{recom}).

\section{Ionic Abundances from Collisionally Excited Lines}
\label{cels}

We have derived ionic abundances of N$^+$, O$^+$, O$^{++}$, Ne$^{++}$, S$^+$, 
S$^{++}$, Cl$^{++}$, Cl$^{3+}$, Ar$^{++}$ and Ar$^{3+}$ from CELs, using the {\sc IRAF} package NEBULAR. 
The atomic data for Cl$^+$ are not implemented in NEBULAR, so we have used an old version of the 
five-level atom program of \citet{shawdufour95} \citep[see][for more details]{garciarojasetal04}.
Ionic abundances are listed in Table~\ref{celabun} and correspond to the mean value of the abundances 
derived from all the individual lines of each observed ion, weighted by their relative intensities.

To derive the ionic abundances for \ts $>$ 0.00 we have used the abundances for {\ts}=0.00 and the formulation by 
\citet{peimbert67} and \citet{peimbertcostero69}. These abundances are also shown in Table~\ref{celabun}.

Several {\ffeii} lines have been detected in the spectra of M8 and M17. Unfortunately, most of them are 
severely affected by fluorescence effects \citep{rodriguez99, verneretal00}. One of the optical {\ffeii} lines 
which is less affected by fluorescence effects is the {\ffeii} $\lambda$8617 line, 
but unfortunately it is in one of our observational gaps. Nonetheless, we have measured 
the {\ffeii} $\lambda$7155 line, both in M8 and M17, a line which is not much affected by 
fluorescence effects \citep{verneretal00}. We have derived an estimation of the Fe$^{+}$ abundance 
from this line, assuming that $I(\lambda7155)$/$I(\lambda8616)$ $\sim$ 1 \citep{rodriguez96} and using 
the calculations by \citet{bautistapradhan96} for the emissivities of the {\ffeii} $\lambda$8617 line. 
We find Fe$^+$/H$^+$ $\sim$ 4.1 $\times$ 10$^{-8}$ 
for M8 and Fe$^+$/H$^+$ $\sim$ 1.1 $\times$ 10$^{-8}$ for M17. Nevertheless these results are only an estimation, 
and we have 
marked them with two colons in Table~\ref{celabun} due to their uncertainty.

\begin{table*}[htbp]\centering
\setlength{\tabnotewidth}{\textwidth}
\tablecols{5}
\setlength{\tabcolsep}{4.6\tabcolsep}
\scriptsize
\caption{Ionic abundances from collisionally excited lines\tabnotemark{a}.}
\label{celabun}
\begin{tabular}{lcccc}
\toprule
& \multicolumn{2}{c}{M8}& \multicolumn{2}{c}{M17} \\
\cmidrule{2-5}
Ion & {\ts}=0.000 & {\ts}=0.040$\pm$0.004 & {\ts}=0.000 & {\ts}=0.033$\pm$0.005 \\
\midrule
N$^{+}$	   & 7.50$\pm$0.03  & 7.67$\pm$0.04 & 6.82$\pm$0.10 & 6.94$\pm$0.10 \\
O$^{+}$    & 8.39$\pm$0.06  & 8.58$\pm$0.07 & 7.84$\pm$0.09 & 7.98$\pm$0.09 \\
O$^{++}$   & 7.86$\pm$0.03  & 8.18$\pm$0.07 & 8.41$\pm$0.04 & 8.67$\pm$0.06 \\
Ne$^{++}$  & 6.95$\pm$0.05  & 7.30$\pm$0.07 & 7.64$\pm$0.04 & 7.93$\pm$0.07 \\
S$^{+}$    & 5.93$\pm$0.04  & 6.10$\pm$0.07 & 5.44$\pm$0.05 & 5.56$\pm$0.06 \\
S$^{++}$   & 6.89$\pm$0.03  & 7.25$\pm$0.07 & 6.90$\pm$0.04 & 7.19$\pm$0.06 \\
Cl$^{+}$   & 4.53$\pm$0.04  & 4.66$\pm$0.06 & 3.95$^{+0.09}_{-0.12}$& 4.06$^{+0.09}_{-0.12}$ \\
Cl$^{++}$  & 5.02$\pm$0.04  & 5.32$\pm$0.06 & 5.04$\pm$0.04& 5.29$\pm$0.06  \\
Cl$^{3+}$  & \nodata        & \nodata       & 3.15: 	    & 3.35: 	    \\
Ar$^{++}$  & 6.21$\pm$0.03  & 6.48$\pm$0.05 & 6.35$\pm$0.04 & 6.57$\pm$0.06 \\
Ar$^{3+}$  & 3.69$\pm$0.09  & 4.01$\pm$0.10 & 4.15$^{+0.12}_{-0.18}$& 4.42$^{+0.13}_{-0.18}$ \\
Fe$^{+}$   & 4.61:          & 4.77:         & 4.05:	    & 4.17:         \\    
Fe$^{++}$  & 5.58$\pm$0.04  & 5.91$\pm$0.06 & 5.24$\pm$0.06& 5.51$\pm$0.08  \\
\bottomrule
\tabnotetext{a}{In units of 12+log(X$^m$/H$^+$).}
\end{tabular}
\end{table*}

The calculations for Fe$^{++}$ have been done with a 34 level model-atom that uses collision strengths from 
\citet{zhang96} and the transition probabilities of \citet{quinet96}. We have used {\ffeiii} lines which 
do not seem affected by blends, 14 in the case of M8 and 5 in the case of M17. We have found an average value and 
a standard deviation of Fe$^{++}$/H$^+$ = (3.78 $\pm$ 0.36) $\times$ 10$^{-7}$ for M8 and Fe$^{++}$/H$^+$ = 
(1.73 $\pm$ 0.12) $\times$ 10$^{-7}$ for M17. Adding errors in {\te} and {\elecd} we finally obtain 12 + 
log(Fe$^{++}$/H$^+$) = 5.58 $\pm$ 0.04 and 5.24 $\pm$ 0.06 for M8 and M17 respectively. The Fe$^{++}$ abundances 
are also 
included in Table~\ref{celabun}.

\section{Ionic Abundances of Heavy Elements from Recombination Lines}
\label{recom}

EPTGR performed a detailed analysis of the excitation mechanisms of permitted heavy element lines in M8. 
In this work we have measured a large number of permitted heavy element lines, but following the study of 
EPTGR we have focused on the lines which are excited purely by recombination. 
Nevertheless, we are going to comment briefly on the N$^{++}$/H$^+$ ratio in both nebulae.

In Table~\ref{nii} we show the N$^{++}$/H$^+$ ratios obtained from permitted lines in M8 and M17. 
\citet{grandi76} argued that resonance fluorescence by the {\hei} $\lambda$508.64 recombination line is the 
dominant mechanism to excite the 4$s^3P^0$ term of {\nii} in the Orion Nebula, and hence, it should 
be responsible for the strengths of multiplets 3 and 5.
Recently, \citet{escalantemorisset05}, 
using tailored photoionization models of the Orion Nebula, estimate that the contribution of recombination 
to the intensity of multiplet 3 (which is one of the less affected by fluorescence effects of those reported in 
this work) 
is about 20 \% of the total intensity of the line. 
Additionally, we have measured a blend of two {\nii} lines of multiplet 19 at $\lambda\lambda$5001.14, 5001.48. 
These lines have upper levels 3d$^3$F$^0_{2,3}$ that are connected to the ground state through weak dipole-allowed 
transitions and could have an important fluorescence contribution \citep{belletal95}; 
\citet{escalantemorisset05} predicted than recombination 
contributes $\sim$43\% to the total intensity of these two lines.
Unfortunately, the only line of this multiplet which is not affected by fluorescence effects is the one at 
$\lambda$5005.15, which is blended with the {\foiii} $\lambda$5007 line.
To test these theoretical predictions we have computed the N$^{++}$ abundance from the line of multiplet 19, 
taking into account the contribution by fluorescence predicted by \citet{escalantemorisset05}, 
and compared it with the N$^{++}$ abundance estimated from the N$^+$ abundances assuming CELs with 
{\ts} $>$ 0.00 and the ionization correction factor for N. 
Also, we have proceeded in the same way with multiplet 3, taking into account that only 20\% of the 
of the line intensities is due to recombination. 
In Table~\ref{niicomp} we show the results obtained for M8 and M17 (this work), NGC~3576 \citep{garciarojasetal04} 
and the Orion Nebula \citep{estebanetal04}. 
For the Orion Nebula and NGC~3576, we have considered also 3d--4f and singlet transitions, which cannot be excited 
by 
resonant fluorescence \citep[see][]{grandi76, escalantemorisset05}. 
In principle, there is better agreement among the abundances obtained from these lines taking into account the 
considerations by 
\citet{escalantemorisset05}; nevertheless there are some puzzling results: the only 3d--4f transition detected 
in NGC~3576 shows the larger deviation from the rest of the values, however, \citet{escalantemorisset05} proposed 
that 
there can be another mechanisms responsible for the enhancement of the intensity of these transitions, so we have 
to consider 
the abundances derived from these lines as high limits; also, from the comparison 
between the recombination  N$^{++}$ abundances and the values obtained from N$^+$/H$^+$ (CELs), the ICF and {\ts} 
in Table~\ref{niicomp} it can be seen that the agreement in M8, M17 and NGC~3576 is not very good, and that in 
Orion is rather poor.  
Nevertheless, the CELs N$^{++}$/H$^+$ ratio is very sensitive to the adopted ICF scheme, and could be reduced as 
much as a factor 
of 2 if the adopted ICF scheme would have been the one by \citet{peimbertcostero69}.
It is clear that the measurement of pure N$^{++}$ recombination lines (i. e. singlet transitions) could be very 
useful to constraint the 
temperature fluctuations scenario, and that much work should be done in this sense, but it is beyond the scope of 
this paper. 

\begin{table*}[htbp]\centering
\setlength{\tabnotewidth}{\textwidth}
\tablecols{8}
\setlength{\tabcolsep}{2.3\tabcolsep}
\scriptsize
\caption{N$^{++}$/H$^+$ ratio from {\nii} permitted lines\tabnotemark{a}}
\label{nii}
\begin{tabular}{cccccccc}
\toprule
& & \multicolumn{3}{c}{M8}& \multicolumn{3}{c}{M17} \\
\cmidrule{3-8}
& & $I$($\lambda$)/$I$(H$\beta$) & \multicolumn{2}{c}{N$^{++}$/H$^+$ ($\times$10$^{-5}$)\tabnotemark{b}}& 
$I$($\lambda$)/$I$(H$\beta$) & \multicolumn{2}{c}{N$^{++}$/H$^+$ ($\times$10$^{-5}$)\tabnotemark{b}} \\
Mult. & $\lambda_0$   & ($\times$10$^{-2}$)   & A 	      & B		& ($\times$10$^{-2}$) & A 	      & B	      
\\
\midrule
3     & 5666.64       & 0.027$\pm$0.004       & 16$\pm$2      & 13$\pm$2       & 0.038$\pm$0.007      & 22$\pm$4      
& 18$\pm$3      \\
      & 5676.02       & 0.015$\pm$0.004       & 18$\pm$5      & 15$\pm$4       & \nodata	      & \nodata       & 
\nodata       \\
      & 5679.56       & 0.034$\pm$0.004       & 10$\pm$1      & 8 $\pm$1       & 0.078$\pm$0.009      & 23$\pm$3      
& 19$\pm$2      \\
      & 5686.21       & 0.009$\pm$0.003       & 17$\pm$6      & 14$\pm$5       & \nodata	      & \nodata       & 
\nodata       \\
      & 5710.76       & 0.010$\pm$0.003       & 17$\pm$5      & 14$\pm$4       & 0.012$\pm$0.005      & 21$\pm$8      
& 17$\pm$7      \\
      & Sum	      & 		      & 13$\pm$1      & 11$\pm$1       &		      & 23$\pm$2      & 19$\pm$2      
\\
5     & 4621.39       & 0.022$\pm$0.004       & 244$\pm$54    & 40$\pm$9       & 0.019: 	      & 216:	      & 
35:	      \\
      & 4630.54       & 0.028$\pm$0.005       & 77$\pm$12     & 13$\pm$2       & 0.055$\pm$0.015      & 154$\pm$43    
& 25$\pm$7      \\
      & 4643.06       & 0.018$\pm$0.004       & 140$\pm$32    & 23$\pm$5       & 0.022: 	      & 175:	      & 
28:	      \\
      & Sum	      & 		      & 116$\pm$11    & 19$\pm$2       &		      & 154$\pm$43    & 25$\pm$7      
\\
19    & 5001.3        & 0.037$\pm$0.009       & 8$\pm$2       & 8$\pm$2        & \nodata	      & \nodata       & 
\nodata       \\
20    & 4788.13       & 0.014$\pm$0.004       & 1447$\pm$391  & 28$\pm$8       & \nodata	      & \nodata       & 
\nodata       \\
      & 4803.29       & 0.011$\pm$0.004       & 642$\pm$214   & 13$\pm$4       & \nodata	      & \nodata       & 
\nodata       \\
      & Sum	      & 		      & 767$\pm$188   & 15$\pm$4       &		      & \nodata       & \nodata       
\\ 
24    & 4994.37       & 0.020$\pm$0.005       & 700$\pm$200   & 30$\pm$10      & \nodata	      & \nodata       & 
\nodata       \\
28    & 5927.82       & 0.014$\pm$0.004       & 3892$\pm$973  & 46$\pm$12      & \nodata	      & \nodata       & 
\nodata       \\
      & 5931.79       & 0.020$\pm$0.004       & 2513$\pm$452  & 30$\pm$5       & 0.031$\pm$0.006      & 
3889$\pm$778  & 46$\pm$9      \\
      & Sum	      & 		      & 2946$\pm$410  & 35$\pm$5       &		      & 3893$\pm$778  & 46$\pm$9      
\\ 
\bottomrule
\tabnotetext{a}{Only lines with intensity uncertainties lower than 40\% have been considered.}
\tabnotetext{b}{Recombination coefficients from \citet{kisieliusstorey02} for cases A and B.}
\end{tabular}
\end{table*}

\begin{table*}[htbp]\centering
\setlength{\tabnotewidth}{\columnwidth}
\tablecols{5}
\setlength{\tabcolsep}{2.6\tabcolsep}
\scriptsize
\caption{Comparison of N$^{++}$/H$^+$ ratios from {\nii} permitted lines.}
\label{niicomp}
\begin{tabular}{ccccc}
\toprule
&  \multicolumn{4}{c}{N$^{++}$/H$^+$ ($\times$10$^{-5}$)\tabnotemark{a}} \\
\cmidrule{2-5}
Mult. & M8 & M17 & Orion & NGC~3576 \\ 
\midrule
3     	& 2 		& 8		& 2	& 2	\\
19    	& 3		& \nodata	& 3	& 4	\\
3d--4f	& \nodata	& \nodata	& $\le$4:& $\le$8\\
singlets& \nodata	& \nodata	& 3:	& 7:	\\
\midrule
CELs\tabnotemark{b}  & 4	& 7		& 6	& 4\\
\bottomrule
\tabnotetext{a}{M8 and M17: this work; Orion Nebula: \citet{estebanetal04}; NGC~3576: \citet{garciarojasetal04}. 
The two colons indicate uncertainties larger than 40\%}
\tabnotetext{b}{N$^{++}$ abundance obtained assuming N/H = N$^{+}$/H$^{+}$ + N$^{++}$/H$^{++}$, where N/H and 
N$^{+}$/H$^{+}$ where obtained 
from CELs and assuming {\ts} $>$ 0.00.}
\end{tabular}
\end{table*}

We have measured 16 permitted lines of {\cii} in the spectrum of M8 and 13 in the spectrum of M17. 
Some of these lines 
(those of multiplets 6, 16.04, 17.02 and 17.04) are $3d-4f$ transitions and are, 
in principle, excited by 
pure recombination \citep[see][]{grandi76}. In these transitions, the abundances 
obtained are case-independent, so 
we have adopted the mean of the values obtained for these transitions as our final 
adopted C$^{++}$/H$^+$ ratio.  
The result for the case-sensitive multiplet 3 gives a C$^{++}$ abundance for case B 
which is rather consistent with the one
adopted here. We have used the effective recombination coefficients computed by 
\citet{daveyetal00} for the abundance calculations. The dispersion of the 
abundances obtained by the different lines is very small, except in the case of {\cii} 
$\lambda$9903.43 line in M17, 
whose intensity seems to be affected by an unknown feature. The final results are in 
excellent agreement with those obtained by EPTGR for M8 (C$^{++}$/H$^+$ = 1.9 $\times$ 10$^{-4}$) 
and by EPTG and 
\citet{tsamisetal03} for M17 (C$^{++}$/H$^+$ = 4.9 $\times$ 10$^{-4}$ and 4.4 
$\times$ 10$^{-4}$, respectively). The complete set of derived individual C$^{++}$/H$^+$ ratios as 
well as the adopted one are shown in Table~\ref{cii}.

\begin{table*}[htbp]\centering
\setlength{\tabnotewidth}{\textwidth}
\tablecols{8}
\setlength{\tabcolsep}{2.0\tabcolsep}
\scriptsize
\caption{C$^{++}$/H$^+$ ratio from {\cii} recombination lines}
\label{cii}
\begin{tabular}{cccccccc}
\toprule
& & \multicolumn{3}{c}{M8} & \multicolumn{3}{c}{M17} \\
\cmidrule{3-8}
& & $I$($\lambda$)/$I$(H$\beta$) & \multicolumn{2}{c}{C$^{++}$/H$^+$ ($\times$10$^{-5}$)\tabnotemark{a}}& 
$I$($\lambda$)/$I$(H$\beta$) & \multicolumn{2}{c}{C$^{++}$/H$^+$ ($\times$10$^{-5}$)\tabnotemark{a}} \\
Mult. & $\lambda_0$ & ($\times$10$^{-2}$) & A & B & ($\times$10$^{-2}$) & A & B\\
\midrule
2    	& 6578.05	& 0.262 $\pm$ 0.008\tabnotemark{b}& 300 $\pm$ 9	& 50 $\pm$ 2 	& 0.358 $\pm$  0.018	& 408 
$\pm$ 20	& 69 $\pm$ 3	\\ 
3    	& 7231.12	& 0.074 $\pm$ 0.004	& 1241 $\pm$ 67 & 18 $\pm$ 1	& 0.129 $\pm$ 0.009	& 2162 $\pm$ 151& 31 
$\pm$ 2	\\
     	& 7236.19	& 0.112 $\pm$ 0.004	& 1045 $\pm$ 37	& 15 $\pm$ 1 	& 0.193 $\pm$ 0.012	& 1800 $\pm$ 112& 26 
$\pm$ 2 	\\
     	& Sum		& 			& 1115 $\pm$ 32	& 16 $\pm$ 1 	& 			& 1929 $\pm$ 90 & 27 $\pm$ 1  	\\ 
4    	& 3918.98	& 0.062 $\pm$ 0.007	& 1210 $\pm$ 137& 385 $\pm$ 43 	& 0.043:		& 820:		& 260: 		\\
     	& 3920.68	& 0.133 $\pm$ 0.008	& 1290 $\pm$ 78	& 410 $\pm$ 25 	& 0.086 $\pm$ 0.023	& 826 $\pm$ 223 & 264 
$\pm$ 71  \\
	& Sum		& 			& 1260 $\pm$ 68	& 400 $\pm$ 22 	& 			& 825 $\pm$ 200	& 263 $\pm$ 62	\\ 
6	& 4267.26	& 0.222 $\pm$ 0.009	& 20 $\pm$ 1	& {\bf 20 $\pm$ 1}& 0.580 $\pm$ 0.035	& 54 $\pm$ 3	& {\bf 
53 $\pm$ 3} \\
16.04	& 6151.43	& 0.009 $\pm$ 0.003	& {\bf 21 $\pm$ 7}& ... 	& 0.018 $\pm$ 0.005	& {\bf 41 $\pm$ 11}& ... 	
\\ 
17.02	& 9903.43	& 0.048 $\pm$ 0.003	& {\bf 18 $\pm$ 1}& ... 	& 0.196 $\pm$ 0.016\tabnotemark{c}	& {\bf 68 
$\pm$ 3}& ... 	\\ 
17.04	& 6461.95	& 0.025 $\pm$ 0.004	& {\bf 22 $\pm$ 4}& ... 	& 0.050 $\pm$ 0.007	& {\bf 44 $\pm$ 6}& ... 	
\\
17.06	& 5342.38	& 0.011 $\pm$ 0.004	& {\bf 19 $\pm$ 7}& ... 	& ...			& ... 	& ... 	\\ 
\midrule
	& Adopted	& 			&\multicolumn{2}{c}{{\bf 20 $\pm$ 1 }}& &\multicolumn{2}{c}{{\bf 48 $\pm$ 3 }} \\ 
\bottomrule
\tabnotetext{a}{Recombination coefficients from \citet{daveyetal00} for cases A and B.}
\tabnotetext{b}{Affected by telluric emission lines.}
\tabnotetext{c}{Blend with an unidentified line.}
\end{tabular}
\end{table*}

The O$^+$ abundance was derived from the {\oi} $\lambda$7771.94 line, the only line of 
multiplet 1 that is not severely affected by 
telluric lines. This multiplet is case independent and is produced mainly by 
recombination because it corresponds 
to a quintuplet transition, and the ground level is a triplet. We also have computed 
the O$^+$/H$^+$ ratio from 
the {\oi} $\lambda$8446.48 line of the multiplet 4, but \citet{grandi75a} showed that 
starlight may contribute significantly 
to the observed strength of the line, which is supported by the fact that the O$^+$/H$^+$ 
ratio implied by this line is between one and 
two orders of magnitude larger.
The effective recombination coefficients were obtained from two sources: \citet{pequignotetal91} and 
\citet{escalantevictor92}. 
Though the results are very similar, we adopted the mean of the abundances obtained with the two 
different coefficients. 
Our results are presented in Table~\ref{oi}. 
The O$^+$ abundance that we have obtained for M8 is 
larger by a factor of 2 than that obtained by EPTGR; whereas, 
as pointed out below, the O$^{++}$ abundance derived from RLs is almost coincident 
in the two works, leading us to propose that the abundance of O$^+$ derived from the 
{\oi} $\lambda$7771.96 line by EPTGR was underestimated by a factor of 2, because of the lower  
spectral resolution and signal-to-noise ratio of their data. The O$^+$ abundance obtained for M17 from the 
$\lambda$7771.94 
line is very uncertain because it is partially blended with a strong sky emission line.

\begin{table*}[htbp]\centering
\setlength{\tabnotewidth}{\textwidth}
\tablecols{8}
\setlength{\tabcolsep}{0.7\tabcolsep}
\scriptsize
\caption{O$^{+}$/H$^+$ ratio from {\oi} permitted lines}
\label{oi}
\begin{tabular}{cccccccc}
\toprule
& & \multicolumn{3}{c}{M8}& \multicolumn{3}{c}{M17} \\
\cmidrule{3-8}
& & $I$($\lambda$)/$I$(H$\beta$) & \multicolumn{2}{c}{O$^{+}$/H$^+$ ($\times$10$^{-5}$)\tabnotemark{a}}& 
$I$($\lambda$)/$I$(H$\beta$) & \multicolumn{2}{c}{O$^{+}$/H$^+$ ($\times$10$^{-5}$)\tabnotemark{a}}\\
Mult. & $\lambda_0$ & ($\times$10$^{-2}$) & A & B & ($\times$10$^{-2}$) & A & B \\
\midrule
1	& 7771.94	& 0.029 $\pm$ 0.003	& 39 $\pm$ 4/30 $\pm$ 3	& \nodata	& 0.025 $\pm$ 0.005\tabnotemark{b} & 33 
$\pm$ 7/25 $\pm$ 5	&  \nodata 	\\
4	& 8446.48\tabnotemark{b}& 0.433 $\pm$ 0.017	& 2454 $\pm$ 96/ 1657 $\pm$ 6	& 493 $\pm$ 9/372 $\pm$ 15 & 0.156 
$\pm$ 0.011	& 890 $\pm$ 62/593 $\pm$ 42 & 179 $\pm$ 13/134 $\pm$ 9 \\ 
\midrule
	& Adopted	& 		&\multicolumn{2}{c}{{\bf 34 $\pm$ 5 }} & &\multicolumn{2}{c}{29 $\pm$ 6 }  \\ 
\bottomrule
\tabnotetext{a}{Recombination coefficients from \citet{pequignotetal91}/\citet{escalantevictor92} for cases A and 
B.}
\tabnotetext{b}{Blended with telluric emission lines.}
\end{tabular}
\end{table*}

\begin{figure}[htbp]
\begin{center}
\includegraphics[width=\columnwidth]{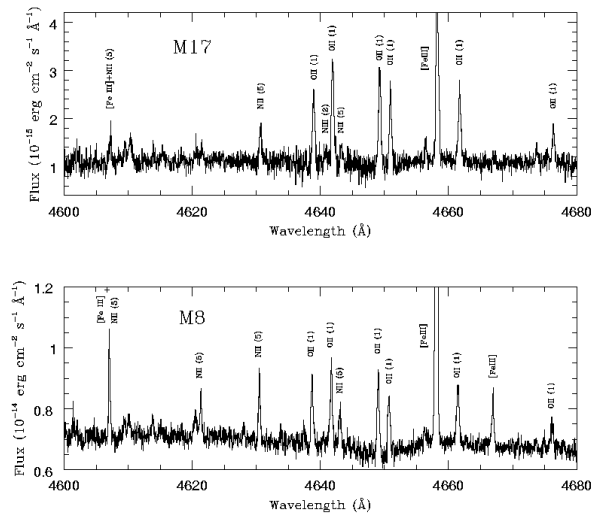}
\caption{Sections of the echelle spectrum of M8 and M17 showing all the lines of multiplet 1 of 
\ion{O}{2}.}
\label{m1oii}
\end{center}
\end{figure} 

We have detected several {\oii} lines in our data. Our spectra of M8 and M17 
present significantly higher signal-to-noise 
than that published before by EPTGR and EPTG, and 
the number of lines to derive the O$^{++}$/H$^+$ ratio has increased. The lower uncertainties and the 
resemblance in the abundances obtained from the different lines make our {\oii} recombination spectra more reliable 
than those of EPTG and EPTGR.   
Figure~\ref{m1oii} shows the high quality of the spectrum in the zone of multiplet 1 of {\oii}. 
This figure can be compared with Figure 1 of \citet{garciarojasetal98}, which shows 
the same spectral zone, and a direct comparison of 
the quality of the spectra can be made. O$^{++}$/H$^+$ ionic abundance ratios are 
presented in Table~\ref{oii}. We have used the same atomic data than in 
\citet{garciarojasetal04} and we have corrected 
for the departure of the local thermodynamic 
equilibrium (LTE) of the upper levels of the transitions of multiplet 1 of {\oii} 
for densities {\elecd} $<$ 10000 cm$^{-3}$, which was pointed out by \citet{tsamisetal03} and \citet{ruizetal03}. 
\citet{apeimbertetal05} proposed an empirical formulation to re-calculate those populations. 
We have applied this NLTE correction
to our data, and we have obtained a very good agreement between the abundances derived 
from individual lines of multiplet 1 and the abundance derived using the sum of all the lines, 
which is expected not to be 
affected by this effect, and with abundances derived from multiplets 2 and 10, which are 
almost case independent. 
The only detection of a $3d-4f$ transition --which are insensitive to optical depths 
effects-- in the spectrum of M8, is a line that is blended 
with a {\cii} line, so its intensity is not reliable.
Therefore, we have adopted as representative of the O$^{++}$/H$^+$ ratio the average 
of the values given by multiplets 1, 2 and 10. 
The O$^{++}$/H$^+$ ratio that we have obtained here for M8 is in very good agreement with 
the one obtained by EPTGR (O$^{++}$/H$^+$ = 2.1 $\times$ 10$^{-4}$); 
on the other hand, for M17, the O$^{++}$ abundance derived here is somewhat lower 
than those 
obtained by EPTGR (O$^{++}$/H$^+$ = 5.5 $\times$ 10$^{-4}$) and 
\citet{tsamisetal03} (O$^{++}$/H$^+$ = 5.7 $\times$ 10$^{-4}$), but this is probably due to 
the different slit size (in the case of EPTG) or to the different slit position (in the case of 
\citealt{tsamisetal03}, 
whose slit is centered about 1' South of our slit center). 

\begin{table*}[htbp]\centering
\setlength{\tabnotewidth}{\textwidth}
\tablecols{6}
\setlength{\tabcolsep}{2.2\tabcolsep}
\scriptsize
\caption{O$^{++}$/H$^+$ ratio from {\oii} recombination lines\tabnotemark{a}}
\label{oii}
\begin{tabular}{cccccc}
\toprule
& & $I$($\lambda$)/$I$(H$\beta$) & \multicolumn{3}{c}{O$^{++}$/H$^+$ ($\times$10$^{-5}$)} \\
Mult. & $\lambda_0$ & ($\times$10$^{-2}$) & A & B & C \\
\midrule
\multicolumn{6}{c}{M8} \\
\midrule
1\tabnotemark{b}& 4638.85	& 0.034 $\pm$ 0.005 & 31 $\pm$ 4/21 $\pm$ 3 & 30 $\pm$ 4/20 $\pm$ 3	& \nodata 	\\
	& 4641.81	& 0.043 $\pm$ 0.005 & 16 $\pm$ 2/18 $\pm$ 2 & 16 $\pm$ 2/17 $\pm$ 2	& \nodata 	\\
	& 4649.14	& 0.041 $\pm$ 0.005 & 9 $\pm$ 1/13 $\pm$ 2  & 8 $\pm$ 1/13 $\pm$ 2	& \nodata 	\\
	& 4650.84	& 0.032 $\pm$ 0.005 & 31 $\pm$ 5/19 $\pm$ 2 & 30 $\pm$ 5/18 $\pm$ 2	& \nodata 	\\
	& 4661.64	& 0.036 $\pm$ 0.005 & 29 $\pm$ 4/20 $\pm$ 3 & 28 $\pm$ 4/19 $\pm$ 2	& \nodata 	\\
	& 4673.73	& \nodata	    & \nodata	 & \nodata	& \nodata			\\
	& 4676.24	& 0.016 $\pm$ 0.004 & 18 $\pm$ 5/19 $\pm$ 4 & 17 $\pm$ 5/19 $\pm$ 4	& \nodata 	\\
	& Sum		& 		    & 18 $\pm$ 1 & {\bf 17 $\pm$ 1}&  \nodata 				\\ 
2	& 4317.14	& 0.011 $\pm$ 0.004 & 22 $\pm$ 8 & 16 $\pm$ 6	& \nodata 	     \\
	& 4319.55	& 0.008: 	    & 15:	 & 11:		& \nodata  	     \\
	& 4345.56\tabnotemark{c}&0.022 $\pm$ 0.005 & 41 $\pm$ 9& 29 $\pm$ 6	& \nodata 	     \\
	& 4349.43	& 0.020 $\pm$ 0.005 & 14 $\pm$ 3 & 10 $\pm$ 2	& \nodata 	     \\
	& 4366.89	& 0.015 $\pm$ 0.004 & 25 $\pm$ 7 & 18 $\pm$ 5	& \nodata 	     \\
	& Sum	 	& 		    & 19 $\pm$ 3 & {\bf 13 $\pm$ 1}&  \nodata 	     \\ \
10\tabnotemark{d}& 4069.62	& 0.067 $\pm$ 0.007 & 26 $\pm$ 3/26 $\pm$ 3	& \nodata 	& \nodata 		\\
	& 4069.89	& 		    & 		 & 		& 					\\
	& 4072.15	& 0.032 $\pm$ 0.006 & 13 $\pm$ 2/13 $\pm$ 2	& \nodata 	& \nodata 		\\ 
	& 4075.86	& \nodata	    & \nodata	& \nodata		& \nodata			\\
	& Sum		&		    & 20 $\pm$ 1/{\bf 20 $\pm$ 1}& \nodata &  \nodata 			\\ 
15\tabnotemark{e}& 4590.97	& 0.005:	    & 29:	& 29:		& \nodata 				\\
	& Sum		& 		    & 29:	& 29:		& \nodata  				\\ 
19\tabnotemark{d}& 4121.48 	& 0.012 $\pm$ 0.005 & 1002 $\pm$ 391 / 746 $\pm$ 291 & 38 $\pm$ 15/ 42 $\pm$ 16 & 38 
$\pm$ 15 / 40 $\pm$ 15 \\
	& 4132.80 	& 0.014 $\pm$ 0.005 & 626 $\pm$ 220/474 $\pm$ 166  & 24 $\pm$ 8/25 $\pm$ 9& 24 $\pm$ 8/23 $\pm$ 8 
\\
	& 4153.30	& 0.028 $\pm$ 0.005 & 927 $\pm$ 176/810 $\pm$ 154  & 35 $\pm$ 7/35 $\pm$ 7& 35 $\pm$ 7/33 $\pm$ 6 
\\
	& Sum		& 		    & 844 $\pm$ 143/678 $\pm$ 115  & 32 $\pm$ 5/33 $\pm$ 6& 32 $\pm$ 5/31 $\pm$ 5 \\ 
3d-4f	&4491.23\tabnotemark{c}& 0.011 $\pm$ 0.004& \nodata	& 70 $\pm$ 26	& \nodata \\ 
\midrule
	& Adopted	& 		&\multicolumn{3}{c}{\bf 17 $\pm$ 1 } 	\\				
\midrule
\multicolumn{6}{c}{M17} \\
\midrule
1\tabnotemark{b}& 4638.85	&    0.093 $\pm$ 0.018     & 85 $\pm$ 16/49 $\pm$ 9	   & 82 $\pm$ 16/47 $\pm$ 9   & 
\nodata    \\
	& 4641.81	&    0.128 $\pm$ 0.019     & 48 $\pm$ 7/56 $\pm$ 8	   & 47 $\pm$ 7/54 $\pm$ 8   & \nodata     \\ 
	& 4649.14	&    0.123 $\pm$ 0.018     & 26 $\pm$ 4/57 $\pm$ 9	   & 25 $\pm$ 4/55 $\pm$ 8   & \nodata     \\
	& 4650.84	&    0.100 $\pm$ 0.018     & 98 $\pm$ 18/48 $\pm$ 9	   & 95 $\pm$ 17/46 $\pm$ 8   & \nodata    \\
	& 4661.64	&    0.119 $\pm$ 0.018     & 97 $\pm$ 15/55 $\pm$ 8	   & 94 $\pm$ 14/54 $\pm$ 8   & \nodata    \\
	& 4673.73	&    0.022:		   & 116:/57:		   & 112:/55:		  & \nodata	   \\
	& 4676.24	&    0.044 $\pm$ 0.014     & 48 $\pm$ 15/55 $\pm$ 18   & 46 $\pm$ 15/53 $\pm$ 17   & \nodata	   
\\
	& Sum		&    			   & 53 $\pm$ 4      & {\bf 51 $\pm$ 4}& \nodata     \\ 
2	& 4317.14	&    0.061 $\pm$ 0.018	   & 119 $\pm$ 36     & 84 $\pm$ 25   & \nodata	    \\
	& 4319.55	&    0.037:	      	   & 72:	    & 51:	 & \nodata	    \\
	& 4345.56\tabnotemark{c}&  0.088 $\pm$ 0.020	   & 163 $\pm$ 37   & 116 $\pm$ 27  & \nodata	    \\
	& 4349.43	&    0.066 $\pm$ 0.018     & 49 $\pm$ 14    & 35 $\pm$ 10   & \nodata	    \\
	& 4366.89	&    0.030:	           & 48:		    & 34:	    & \nodata	    \\
	& Sum	 	&   		           & 68 $\pm$ 12	    & {\bf 48 $\pm$ 9}& \nodata     \\ 
10\tabnotemark{d}& 4069.62	&    0.190 $\pm$ 0.027     & 75 $\pm$ 11/73 $\pm$ 10& \nodata	   & \nodata	   \\
	& 4069.89	&   			   &	   &	 &  \\
	& 4072.15	&    0.091 $\pm$ 0.022     & 38 $\pm$ 9/38 $\pm$ 9 &  \nodata	   & \nodata	   \\
	& 4075.86	&    0.087 $\pm$ 0.022     & 25 $\pm$ 6/25 $\pm$ 6 &  \nodata	   & \nodata	   \\
	& Sum		&   			   & 44 $\pm$ 5/{\bf 43 $\pm$ 5}& \nodata &  \nodata\\ 
19\tabnotemark{d}& 4121.48 	&    \nodata		   & \nodata	  & \nodata	& \nodata	\\
	& 4132.80 	&    \nodata		 & \nodata	 & \nodata	 & \nodata	 \\
	& 4153.30	&    0.092 $\pm$ 0.021   & 3105 $\pm$ 715/2714 $\pm$ 625& 117 $\pm$ 27/118 $\pm$ 27& 117 $\pm$ 
27/110 $\pm$  25\\
	& Sum		&    			 & 3105 $\pm$ 715/2714 $\pm$ 625& 117 $\pm$ 27/118 $\pm$ 27& 117 $\pm$ 27/110 $\pm$ 
25\\ 
\midrule
	& Adopted	&			   & \multicolumn{3}{c}{\bf 48 $\pm$ 2 } \\ 
\bottomrule
\tabnotetext{a}{Only lines with intensity uncertainties lower than 40 \% have been considered. Recombination 
coefficients 
are those of \citet{storey94} for cases A and B unless otherwise stated.}
\tabnotetext{b}{Not corrected from NLTE effects/corrected form NLTE effects (see text).}
\tabnotetext{c}{Blend.}
\tabnotetext{d}{Values for  LS coupling \citep{storey94}/intermediate coupling \citep{liuetal95}.}
\tabnotetext{e}{Dielectronic recombination rates by \citet{nussbaumerstorey84}.}
\end{tabular}
\end{table*}

\section{Total Abundances}
\label{abuntot}

To derive the total gaseous abundances we have to correct for the unseen ionization stages 
by using a set of ICFs. We have adopted the same scheme used in 
\citet{garciarojasetal05} and \citet{garciarojasetal06}. 

\begin{table*}[htbp]\centering
\setlength{\tabnotewidth}{\textwidth}
\newcommand{\DS}{\hspace{5\tabcolsep}} 
\tablecols{6}
\setlength{\tabcolsep}{3.2\tabcolsep}
\scriptsize
\caption{Total Gaseous Abundances.}
\label{totabun}
\begin{tabular}{lcc l cc}
\toprule
& \multicolumn{2}{c}{M8}&& \multicolumn{2}{c}{M17}  \\
\cmidrule{2-3}
\cmidrule{5-6}
Element & {\ts}=0.000 & {\ts}=0.040$\pm$0.004 && {\ts}=0.000 & {\ts}=0.033$\pm$0.005\\
\midrule
He		& 10.87$\pm$0.01 	& 10.85$\pm$0.01&	& 10.97$\pm$0.01 	& 10.97$\pm$0.01 	\\
C\tabnotemark{a}	& 8.61/8.69$\pm$0.09	& 8.70/8.69$\pm$0.09&    & 8.77$\pm$0.04 	& 8.77$\pm$0.04 	\\
N		& 7.72$\pm$0.03 	& 7.96$\pm$0.06& 	& 7.62$\pm$0.12 	& 7.87$\pm$0.13 	\\
O		& 8.51$\pm$0.05		& 8.73$\pm$0.05& 	& 8.52$\pm$0.04		& 8.76$\pm$0.05 	\\
O\tabnotemark{b}	& 8.71$\pm$0.04		& 8.71$\pm$0.04& 	& 8.76$\pm$0.04 	& 8.76$\pm$0.04 	\\
Ne		& 7.81$\pm$0.12 	& 8.03$\pm$0.13& 	& 7.74$\pm$0.07 	& 8.01$\pm$0.09 	\\
S		& 6.94$\pm$0.03 	& 7.28$\pm$0.06& 	& 7.01$\pm$0.04 	& 7.33$\pm$0.06 	\\
Cl\tabnotemark{c}	& 5.14$\pm$0.04		& 5.41$\pm$0.06& 	& 5.08/5.06$\pm$0.04	& 5.32/5.30$\pm$0.06 	\\
Ar		& 6.52$\pm$0.04 	& 6.69$\pm$0.07& 	& 6.39$\pm$0.14 	& 6.59$\pm$0.15 	\\
Fe		& 5.69$\pm$0.08 	& 6.04$\pm$0.09& 	& 5.87$\pm$0.12		& 6.22$\pm$0.14 	\\
\bottomrule
\tabnotetext{a}{For M8: ICF from a [C II] UV line/ICF from \citet{garnettetal99}.}
\tabnotetext{b}{For M8, O$^{++}$/H$^+$ and O$^+$/H$^+$ from RLs. For M17, O$^{++}$/H$^+$ from RLs and O$^+$/H$^+$ 
from CELs and t$^2$.}
\tabnotetext{c}{For M17: From Cl$^{+}$/H$^+$+Cl$^{++}$/H$^+$+Cl$^{3+}$/H$^+$/Using ICF from 
\citet{peimberttorrespeimbert77}.}
\end{tabular}
\end{table*}

The total abundances for N, O, Ne, S, Cl, Ne and Fe have been derived using CELs and an ICF, for {\ts}=0.00 and 
{\ts}$>$0.00. 
For C we have computed the total abundance from the C$^{++}$ abundance derived from RLs and an ICF derived from 
photoionization 
models by \citet{garnettetal99}; for M8, we have also considered the ICF obtained from the C$^+$/H$^+$ ratio 
obtained from 
$IUE$ observations of the $\lambda$ [\ion{C}{2}] 2326 line
For M8 we have derived also the O/H ratio by adopting O$^{++}$/H$^+$ and O$^+$/H$^+$ from RLs. 
For M17 we have computed the total oxygen abundance, adopting O$^{++}$/H$^+$ from RLs and O$^+$/H$^+$ from CELs and 
{\ts}$>$0.00, 
because O$^+$/H$^+$ from RLs was not reliable (see \S~\ref{recom}).  
In Table~\ref{totabun} we present the adopted total abundances for M8 and M17. 

\section{Detection of Deuterium Balmer lines in M8 and M17}
\label{deuterium}

\begin{figure}[htbp]
\includegraphics[width=\columnwidth]{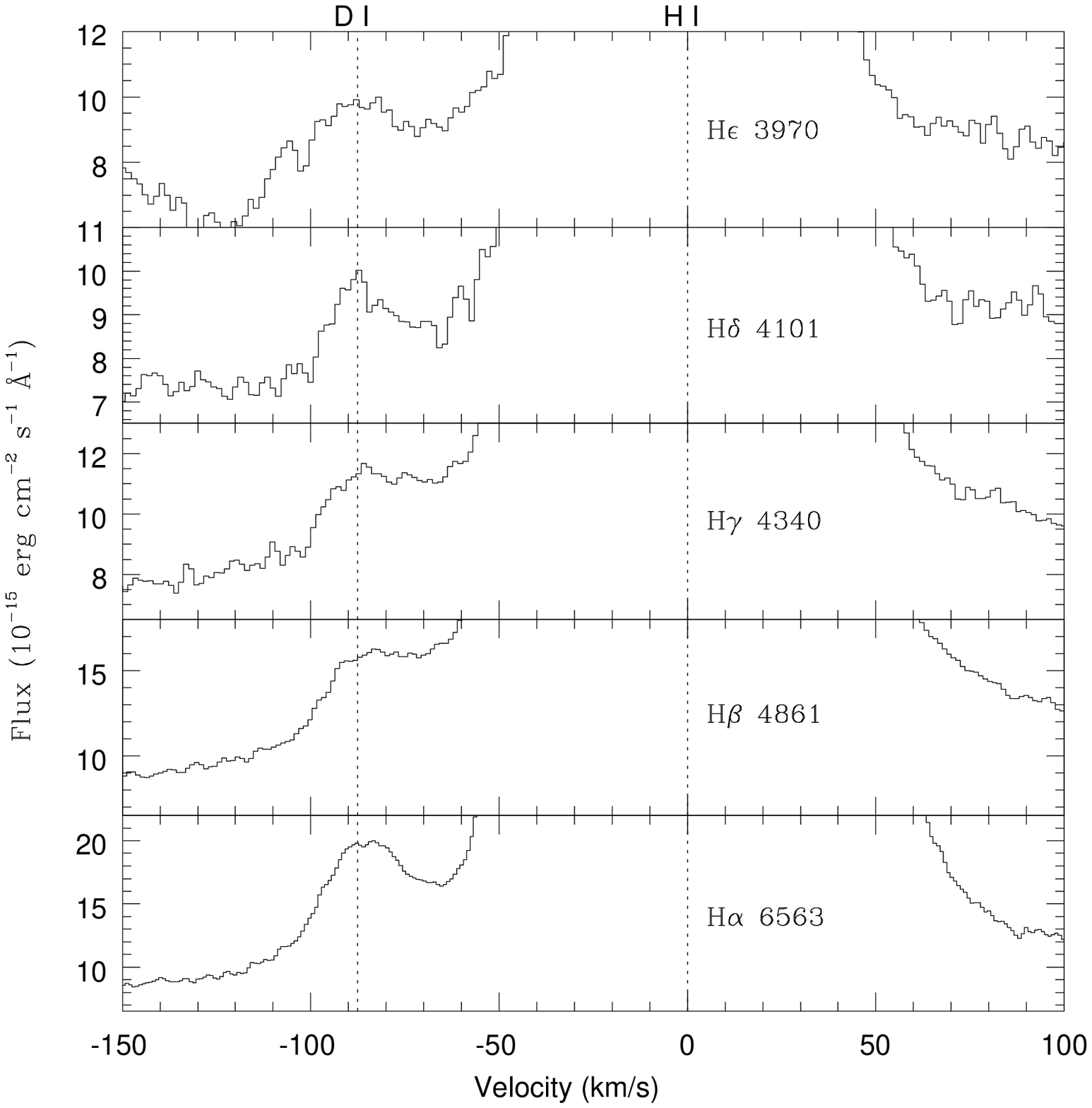}
\caption{Wings of {\ha} to H$\epsilon$ in M8. The {\hi}  lines are centered at 0 km s$^{-1}$  
velocity. 
The dotted line of the left correspond to the average wavelength adopted for the {\di} lines.}
\label{deum8}
\end{figure}

\begin{figure}[htbp]
\includegraphics[width=\columnwidth]{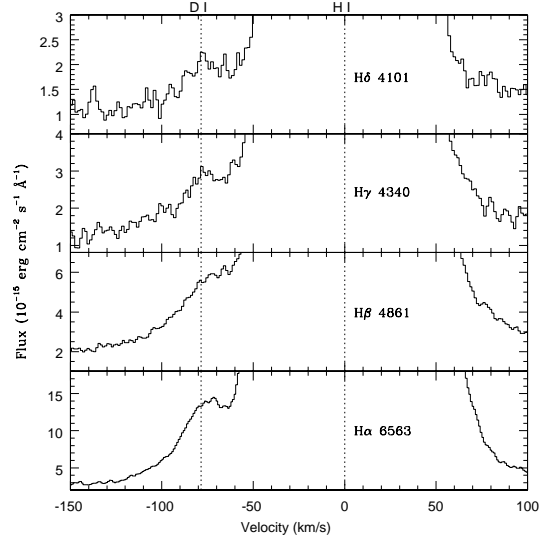}
\caption{Same as figure~\ref{deum8}, for M17. It is not clear that these features could be {\di} lines (see text).}
\label{deum17}
\end{figure}

\citet{hebrardetal00b} reported the detection of deuterium Balmer lines 
in the spectrum of M8, but they did not find these features in M17. 
In M8 these authors 
detected from D$\alpha$ to D$\zeta$. 
We have detected several weak features in the blue wings of {\hi} Balmer lines in M8 --from H$\alpha$ to 
H$\epsilon$-- 
and in M17 --from H$\alpha$ to H$\delta$--
(see figures~\ref{deum8} and \ref{deum17}). The apparent shifts in radial velocity of 
these lines with respect to the {\hi} ones are $-87.6$ km s$^{-1}$ for M8 and $-78.5$ km s$^{-1}$ 
for M17, which are similar to the isotopic shift of deuterium, $-81.6$ km s$^{-1}$.

\begin{figure}[htbp]
\includegraphics[width=\columnwidth]{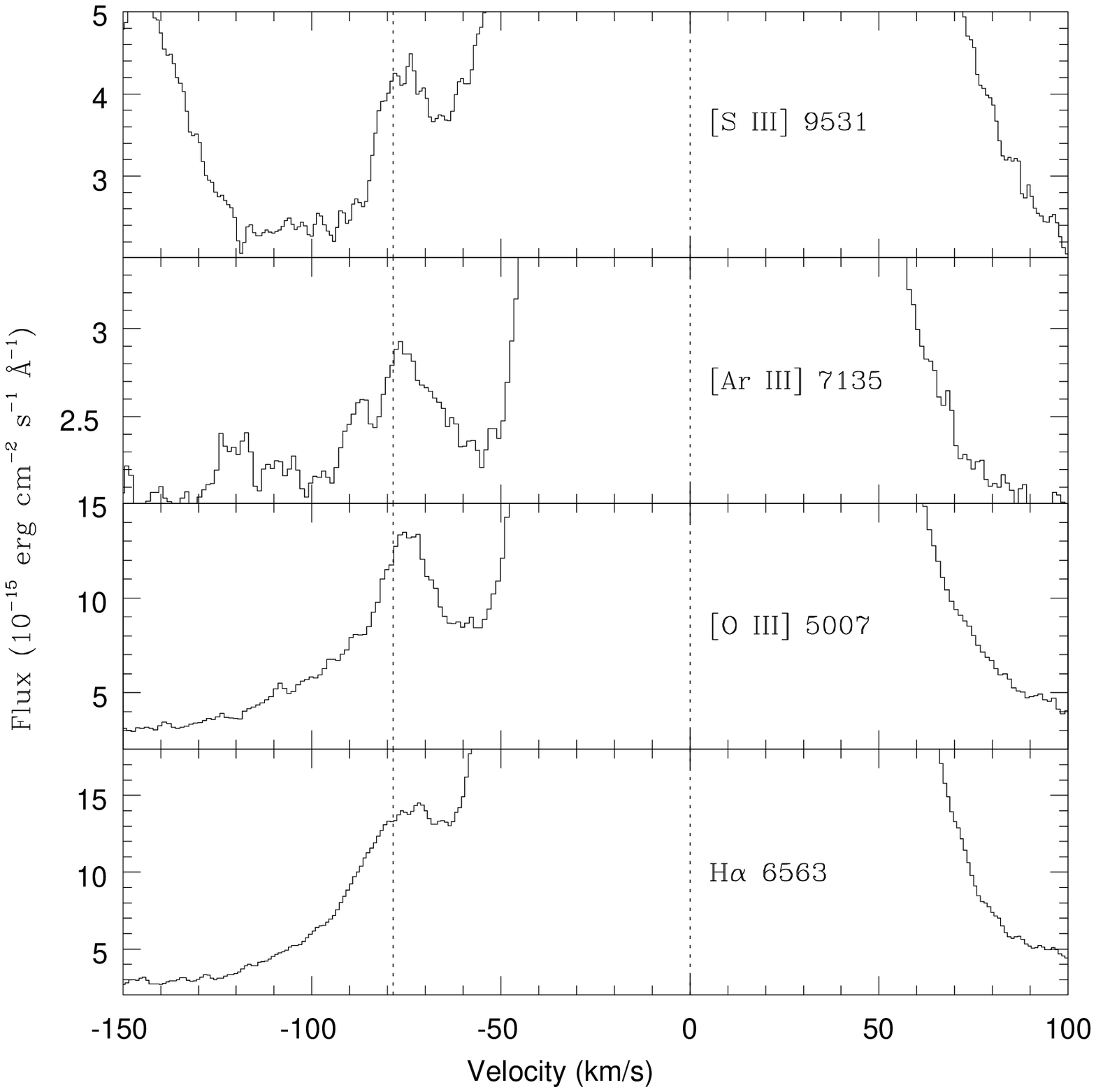}
\includegraphics[width=\columnwidth]{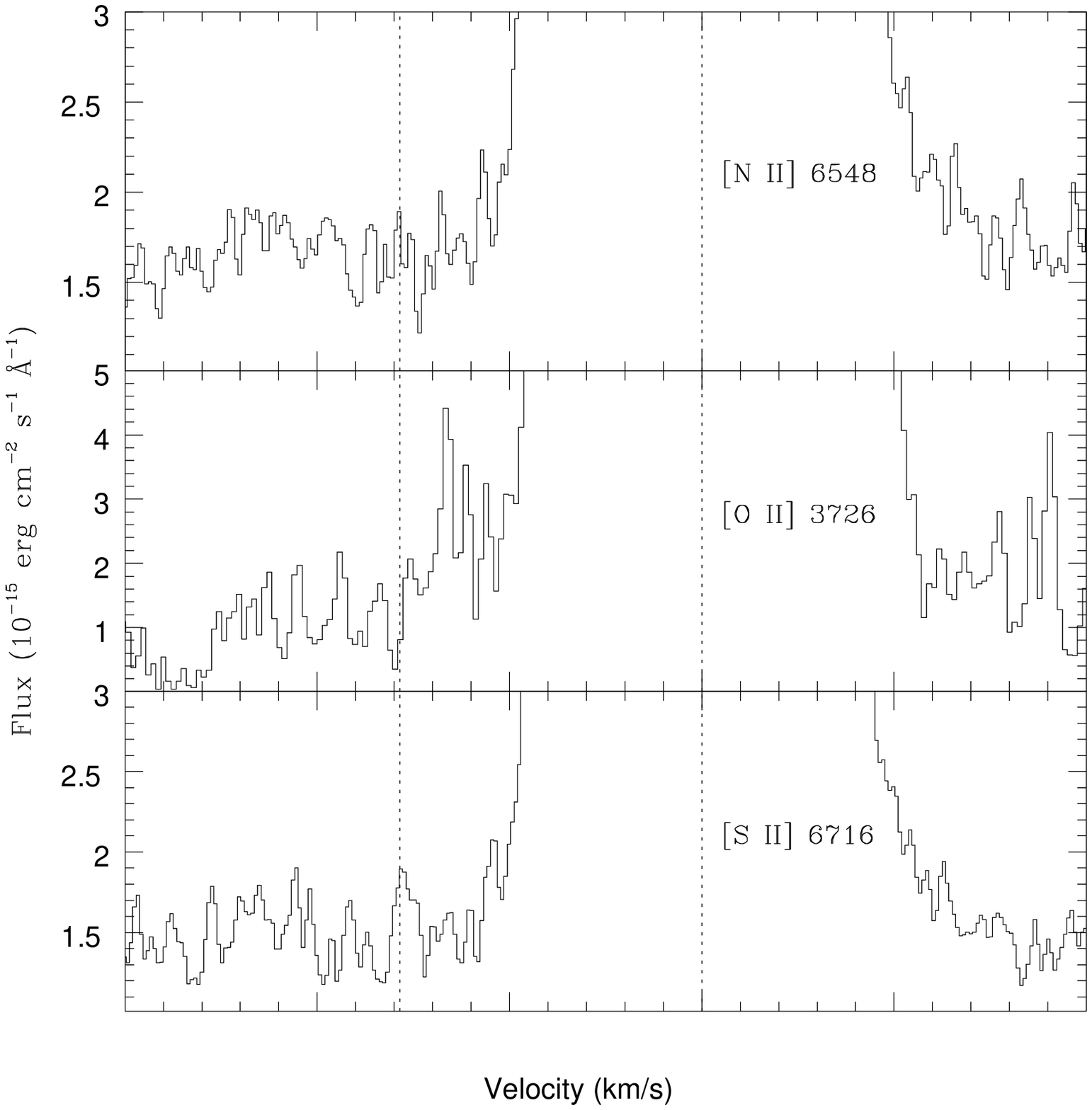}
\caption{Same as figure~\ref{deum8}, for the wings of {\ha}, {\foiii} $\lambda$5007, {\fariii} $\lambda$7135, 
and {\fsiii}$\lambda$9531 (top) and {\foii} $\lambda$3726, {\fsii} $\lambda$6716, 
and {\fnii}$\lambda$6548 (bottom) in M17. The dotted line correspond to the average wavelength adopted for the 
blue-shifted {\hi} lines.}
\label{bluem17}
\end{figure}

For M8, these weak features could be discarded as high-velocity components of hydrogen following the criteria 
established by 
\citet{hebrardetal00b} to identify {\di} lines: a) they are narrower than the {\hi} line, probably because {\di} 
lines arise from much colder material in the photon-dominated region (PDR); b) there are no similar high velocity 
components 
associated to bright lines of other ions. Furthermore, the Balmer decrement of these lines follows closely the 
standard fluorescence 
models by \citet{odelletal01} for the Orion nebula (see Table~\ref{deuchar}), indicating that fluorescence should 
be the main excitation mechanism of 
the {\di} lines. The difference on the apparent shift in radial velocity measured for these lines with respect to 
the isotopic shift of deuterium 
(see above) is probably due to relative motions of the gas in the photon dominated region or PDR --where the 
deuterium Balmer 
lines are supposed to be formed-- with respect to the main emitting layer of the nebula. 
Table~\ref{deuchar} shows the main characteristics of the {\di} Balmer lines in M8. 

\begin{table}[htbp]\centering
\setlength{\tabnotewidth}{\columnwidth}
\tablecols{5}
\setlength{\tabcolsep}{0.7\tabcolsep}
\scriptsize
\caption{Deuterium Balmer line properties in M8.}
\label{deuchar}
\begin{tabular}{ccccc}
\toprule
Line & Velocity & FWHM {\di} & FWHM {\hi} &{\di}/{\hi} ratio \\
     &  shift (km s$^{-1}$) & (km s$^{-1}$)   & (km s$^{-1}$) & ($\times$10$^{-4}$)  \\
\midrule
$\alpha$ & $-87.3$ & $<$ 10: & 24 & 2.9 $\pm$ 0.2  \\
$\beta$  & $-87.6$ & $<$ 10: & 19 & 3.6 $\pm$ 0.5 \\
$\gamma$ & $-87.7$ & $<$ 10: & 19 & 4.1 $\pm$ 1.0 \\
$\delta$ & $-87.7$ & $<$ 10: & 19 & 6.5 $\pm$ 2.0 \\
$\epsilon$ & $-87.6$ & $<$ 10:& 19 & 8.8 $\pm$ 3.0 \\ 
\bottomrule
\end{tabular}
\end{table}

\citet{hebrardetal00b} identified the weak features in the blue wings of {\hi} Balmer lines in M17 as high velocity 
components 
of hydrogen mainly because of the presence of very similar features in the wings of {\fnii}, {\foii} and {\foiii} 
lines.

From our data we do not have a clear cut case because: a) the width of the {\hi} blue-shifted feature (HBSF) is 
narrow like typical Balmer 
{\di} lines; b) we cannot compare the ratios of  HBSF/{\hi} with
the standard fluorescence models of {\di} Balmer lines by \citet{odelletal01}; in principle the ratios seem to fit 
the model, but 
errors are so high that it might be possible for the HBSF/{\hi} ratios to be constant (see 
Table~\ref{bluefeachar}); and c)
two {\foiii} lines --$\lambda$$\lambda$4959, 5007--, two {\fariii} lines --$\lambda$$\lambda$7135, 7751-- 
and two {\fsiii} lines --$\lambda$$\lambda$9069, 9531-- present blue 
counterparts at about $\sim$74, 77 and 78 km $s^{-1}$ respectively, which differ in only a few km s$^{-1}$ from the 
average 
shift of the blue shifted {\hi} lines (see Figure~\ref{bluem17} (top)). These counterparts 
are not detected in the wings of {\fnii}, {\foii} and {\fsii} lines (see Figure~\ref{bluem17} (bottom)).

Table~\ref{bluefeachar} shows the main characteristics of the HBSFs in M17. 

\begin{table}[htbp]\centering
\setlength{\tabnotewidth}{\columnwidth}
\tablecols{5}
\setlength{\tabcolsep}{0.7\tabcolsep}
\scriptsize
\caption{{\hi} blue-shifted feature properties in M17.}
\label{bluefeachar}
\begin{tabular}{ccccc}
\toprule
Line & Velocity & FWHM HBSF & FWHM {\hi} &HBSF/{\hi} ratio \\
     &  shift (km s$^{-1}$) & (km s$^{-1}$)   & (km s$^{-1}$) & ($\times$10$^{-4}$)  \\
\midrule
H$\alpha$ & $-78.6$ & $<$ 10: & 26 & 3.9 $\pm$ 0.4 \\
H$\beta$  & $-77.7$ & $<$ 10: & 25 & 4.2 $\pm$ 1.3 \\
H$\gamma$ & $-78.7$ & $<$ 10: & 25 & 5.3:  \\
H$\delta$ & $-78.9$ & $<$ 10: & 25 & 11.7:   \\ 
\bottomrule
\end{tabular}
\end{table}

From a simple visual inspection of Figure~\ref{bluem17} comparing the width and central wavelength of the blue 
components of H$\alpha$ 
and some forbidden lines, it is possible for the HBSFs to be a blend of {\di} emission and a blushifted 
high-velocity {\hi} component, 
but, with the available constraints, we cannot guarantee it.

\section{Comparison with previous abundance determinations.}
\label{comp}

{From} the comparison of our data of M8--HGS with those published by EPTGR, it seems that a small 
difference in the volume covered by the slit in this zone is sufficient to change significantly the 
ionization degree of the gas; this fact was also pointed out by \citet{sanchezpeimbert91}. 
This is because the emission comes from a range of densities, 
temperatures, degrees of ionization and sometimes extinctions within the column of gas. 
Nonetheless, total abundances should be invariant. In particular, the total oxygen abundance is not affected 
by the uncertainty of using an ICF because all the 
stages of ionization of this element have been detected in our optical spectra. 
In Table~\ref{compa} we show the comparison between total abundances obtained in this work and those obtained in 
previous works for M8 
and M17. Uncertainties reported in previous works for the total abundances are about 0.1 dex or even larger; taking 
into account 
the heterogeneity of the error criteria among the different works we have adopted that errors should be about 
0.1 dex in this work.

The total abundances of M8 are in quite good agreement with those derived by EPTGR, 
within the uncertainties and taking into account that the ICFs for neon and argon 
(which present the largest deviations from our data) are reported as uncertain. Also, the ICF scheme and the atomic 
data used by EPTGR for iron are different to those used here. Making use of our atomic data and ICF 
scheme, the abundances obtained with the EPTGR data lead to a much better agreement (see Table~\ref{compa}). 
We have proceeded in a similar way with the data of \citet{peimbertetal93b} and \citet{rodriguez99, rodriguez99b}, 
reaching the same conclusions. 

\begin{table*}[htbp]
\centering 
\setlength{\tabnotewidth}{\textwidth}
\tablecols{11}
\setlength{\tabcolsep}{2.2\tabcolsep}
\scriptsize
\caption{Comparison with previous determinations\tabnotemark{a}.}
\label{compa}
\begin{tabular}{lcccclccccc}
\toprule
& \multicolumn{4}{c}{M8} && \multicolumn{5}{c}{M17} \\
\cmidrule{2-5}\cmidrule{7-11}
Element & (1) & (2) & (3) & (4)\tabnotemark{c} && (1) & (4)\tabnotemark{d} & (5) & (6) & (7) \\
\midrule
N           & 7.72$\pm$0.03  & 7.68 	& 7.75 	    & 7.60     && 7.62$\pm$0.12&  7.50  & 7.59    & 7.55	   & 
7.57    \\
O	    & 8.51$\pm$0.05  & 8.49 	& 8.54 	    & 8.43     && 8.52$\pm$0.04&  8.53  & 8.51    & 8.51	   & 8.55    
\\
Ne	    & 7.81$\pm$0.12  & 7.76 	& 7.83 	    & \nodata  && 7.74$\pm$0.07& \nodata& 7.81    & 7.78	   & 7.79    
\\
S           & 6.94$\pm$0.03  & 6.96 	& 7.03 	    & 6.95     && 7.01$\pm$0.04&  6.99  & 7.03    & 6.84	   & 
7.05    \\
Cl          & 5.14$\pm$0.04  & 5.21 	& \nodata   & 5.20     && 5.06$\pm$0.04&  5.02  & 5.03    & 5.03	   & 
5.07    \\
Ar          & 6.52$\pm$0.04  & 6.53 	& 6.48 	    & 6.60     && 6.39$\pm$0.14&  6.36  & 6.26    & 6.35	   & 
6.39    \\
Fe          & 5.69$\pm$0.08  & 5.80 	& \nodata   & 5.72     && 5.87$\pm$0.12&  5.75  & \nodata & 5.88	   & 
\nodata \\ 
\bottomrule
\tabnotetext{a}{In units of 12+log(X/H). Abundances have been recomputed using our atomic data and ICF scheme (see 
text).}
\tabnotetext{b}{(1) This work; (2) EPTG; (3) \citet{peimbertetal93b}; (4) \citet{rodriguez99, rodriguez99b}; 
(5) \citet{tsamisetal03}; (6) EPTGR; (7) \citet{peimbertetal92}.}
\tabnotetext{c}{These data are an average of the total abundances obtained in the two slit positions located 
southwards of the 
Hourglass.}
\tabnotetext{d}{These data are an average of the results obtained in three slit positions in M17. }
\end{tabular}
\end{table*}

The abundances of M17 are in good agreement, within the errors, with those derived by EPTG, \citet{peimbertetal92}, 
\citet{rodriguez99b} and \citet{tsamisetal03}. 
The larger differences among the different works are probably due to the different sets of atomic data and ICFs 
used. 
Proceeding in the same way as in M8 we have recomputed the abundances for M17 using our atomic data 
and ICF scheme; in this case the differences with the previous calculations are of the order of 0.1 dex. Therefore, 
we can conclude 
that errors in the line intensities are not responsible for the differences among abundances of these elements by 
different authors 
(both in M8 and M17) 
and we emphasize the high robustness of the abundances determined for these objects, bearing in mind the 
uncertainties due to atomic 
data and ICF schemes (see Table~\ref{compa}). 

\section{SUMMARY}
\label{conclu}

We present new echelle spectroscopy in the 3100--10450  {\rm \AA} range of the Hourglass Nebula in M8 and 
a bright rim of M17.

We have determined the physical conditions of M8 and M17 making use of a large number of diagnostic line ratios.

We have derived ionic abundances from CELs as well as C$^{++}$/H$^+$ and O$^{++}$/H$^+$ ratios making use of 
RLs in these nebulae. 
The ionic abundances obtained from RLs are in very good agreement with those obtained in previous works. 

The very good agreement between our results --that have been obtained making use of state-of-the-art atomic data-- 
and 
the best abundance determinations from the literature for M8 and M17 allow us to assure that the total abundances 
of these 
nebulae are very well established.

We have obtained an average $\ts$ of 0.040 $\pm$ 0.004 for M8 and 0.033 $\pm$ 0.005 for M17 which are rather 
similar to the values 
derived in previous works for these two nebulae. Also, it is remarkable the excellent agreement among the {\ts} 
values obtained through 
independent methods. This behavior is consistent with the temperature fluctuations scenario.

We confirm the detection of deuterium Balmer emission lines in M8 and possibly in M17, although in this case 
there seems to be an accidental contamination of a blueshifted high-velocity {\hi} component.

JGR would like to thank 
the members of the Instituto de Astronom\'ia, UNAM, and of the Instituto Nacional de Astrof\'{\i}sica, \'Optica y 
Electr\'onica, INAOE, for their always warm hospitality. This work has been partially funded by the Spanish 
Ministerio de Ciencia y Tecnolog\'{\i}a (MCyT) under project AYA2001-0436 and AYA2004-07466. MP received partial
support from CONACyT (grant 46904). MR acknowledges support from Mexican CONACyT project 
J37680-E. 
MTR received partial support from FONDAP(15010003) and Fondecyt(1010404).

\setcounter{table}{1}
\begin{table*}[htbp]\centering
\setlength{\tabnotewidth}{0.8\textwidth}
\tablecols{8}
\setlength{\tabcolsep}{2.0\tabcolsep}
\scriptsize
\caption{Observed and reddening-corrected line fluxes with respect to F(H$\beta$)=100.}
\label{lineidm8m17}
\begin{tabular}{llccccrl}
\toprule
$\lambda_{\rm 0}$& & & $\lambda_{\rm obs}$& & & err & Notes \\
({\rm \AA})& Ion& Mult.& ({\rm \AA})& $F(\lambda)$& $I(\lambda)$\tabnotemark{a}& (\%)\tabnotemark{b} & \\
\midrule
\multicolumn{8}{c}{\bf \large{M 8}}\\
\midrule
3187.84 & He I & 3  & 3187.70  &    0.944  & 0.315 &  13  &  \\
3354.55 & He I & 8  & 3354.54  &    0.146  & 0.137 &  19  &  \\
3447.59 & He I & 7  & 3447.53  &    0.181  & 0.222 &  17  &  \\
3487.73 & He I & 42 & 3487.70  &    0.057  & 0.075 &  :   &  \\
3498.66 & He I & 40 & 3498.61  &    0.035  & 0.047 &  :   &  \\
3512.52 & He I & 38 & 3512.48  &    0.096  & 0.128 &  27  &  \\
3530.50 & He I & 36 & 3530.46  &    0.092  & 0.124 &  28  &  \\
3554.42 & He I & 34 & 3554.39  &    0.163  & 0.221 &  17  &  \\
3587.28 & He I & 32 & 3587.24  &    0.181  & 0.242 &  17  &  \\
3613.64 & He I & 6  & 3613.61  &    0.291  & 0.381 &  12  &  \\
3634.25 & He I & 28 & 3634.20  &    0.300  & 0.385 &  12  &  \\
3656.10 & H I & H37 & 3656.08  &    0.036  & 0.045 &  :   &  \\
3657.27 & H I & H36 & 3657.24  &    0.088  & 0.110 &  29  &  \\
3657.92 & H I & H35 & 3657.89  &    0.106  & 0.134 &  25  &  \\
3658.64 & H I & H34 & 3658.60  &    0.111  & 0.139 &  24  &  \\
3659.42 & H I & H33 & 3659.39  &    0.143  & 0.179 &  19  &  \\
3660.28 & H I & H32 & 3660.24  &    0.161  & 0.202 &  17  &  \\
3661.22 & H I & H31 & 3661.18  &    0.196  & 0.246 &  16  &  \\
3662.26 & H I & H30 & 3662.23  &    0.212  & 0.266 &  15  &  \\
3663.40 & H I & H29 & 3663.37  &    0.236  & 0.295 &  14  &  \\
3664.68 & H I & H28 & 3664.62  &    0.271  & 0.338 &  12  &  \\
3666.10 & H I & H27 & 3666.05  &    0.299  & 0.374 &  12  &  \\
3667.68 & H I & H26 & 3667.64  &    0.360  & 0.449 &   11  &  \\
3669.47 & H I 	& H25 	& 3669.42  &    0.386  &   0.481  &   10  &  \\
3671.48 & H I 	& H24 	& 3671.43  &    0.425  &   0.528  &   10  &  \\
3673.76 & H I 	& H23 	& 3673.71  &    0.461  &   0.573  &   10  &  \\
3676.37 & H I 	& H22 	& 3676.32  &    0.505  &   0.627  &   9  &  \\
3679.36 & H I 	& H21 	& 3679.30  &    0.567  &   0.702  &   9  &  \\
3682.81 & H I 	& H20 	& 3682.76  &    0.587  &   0.727  &   9  &  \\
3686.83 & H I 	& H19 	& 3686.79  &    0.652  &   0.805  &   9  &  \\
3691.56 & H I 	& H18 	& 3691.51  &    0.776  &   0.956  &   9  &  \\
3697.15 & H I 	& H17 	& 3697.10  &    0.954  &   1.174  &   8  &  \\
3703.86 & H I 	& H16 	& 3703.80  &    1.065  &   1.308  &   8  &  \\
3705.04 & He I 	& 25 	& 3704.97  &    0.412  &   0.506  &   11  &  \\
3711.97 & H I 	& H15 	& 3711.92  &    1.276  &   1.566  &   8  &  \\
3721.83 &  [S III] & 2F      & 3721.81  &    2.260  &	2.772  &   8  &  \\
3721.94 &  H I     & H14     &          &	    &	       &      &  \\
3726.03 &  [O II]   & 1F      & 3726.01  &  100.407  &  123.200 &   8  &  \\
3728.82 &  [O II]   & 1F      & 3728.77  &   70.305  &   86.287 &   8  &  \\
3734.37 &   H I    & H13     & 3734.32  &    1.908  &  2.343   &   8  &  \\
3750.15 &   H I    & H12     & 3750.10  &    2.378  &  2.933   &   8  &  \\
3770.63 &   H I    & H11     & 3770.58  &    3.066  &  3.818   &   8  &  \\
3784.89 & He I     & 64      & 3784.88  &    0.019  &  0.023   &  23  &  \\
3797.90 &   H I    & H10     & 3797.85  &    4.040  &  5.132   &   8  &  \\
3805.78 & He I     & 63      & 3805.64  &    0.030  &	0.039  &  17  &  \\
3819.61 &   He I   & 20      & 3819.58  &    0.694  &	0.902  &   8  &  \\
	&   ?	   &	     & 3831.14  &    0.013  &    0.017 &  30  & c \\
3831.66 & S II     &	     & 3831.64  &    0.021  &	 0.027 &  21  & c \\
3833.57 &   He I   & 62      & 3833.50  &    0.025  &	 0.033 &  18  &  \\
3835.39 & H I	   & H9      & 3835.33  &    5.567  &	7.245  &   8  &  \\
3838.37 & N II     & 30      & 3838.23  &    0.019  &	0.025  &  22  &  \\
3853.66 & Si II    & 1       & 3853.61  &    0.011  &	0.015  &  33  &  \\
3856.02 &   Si II  & 1       & 3855.98  &    0.145  &	0.189  &   9  &  \\
3856.13 &   O II   & 12      &	        &	    &	       &      &  \\
3862.59 & Si II    & 1       & 3862.56  &    0.089  &  0.116   &   11  &  \\
3867.48 & He I     & 20      & 3867.45  &    0.053  &  0.069   &   8  &  \\
\bottomrule
\tabnotetext{a}{For M8, c(H$\beta$)=0.94 and $I$(H$\beta$=$2.543 \times 10^{-11}$ 
ergs cm$^{-2}$ s$^{-1}$). For M17, c(H$\beta$)=1.17 and $I$(H$\beta$=$1.201 \times 10^{-11}$ ergs cm$^{-2}$ 
s$^{-1}$).}
\tabnotetext{b}{Colons indicate uncertainties larger than 40 \%.}
\tabnotetext{c}{Dubious identification.}
\tabnotetext{d}{Full Table available in http://www.iac.es/galeria/jogarcia/PAPERS/table2.pdf}
\end{tabular}
\end{table*}

\end{document}